\documentclass[fleqn,usenatbib]{mnras}

\usepackage{newtxtext,newtxmath}
\usepackage[T1]{fontenc}
\usepackage{ae,aecompl}

\usepackage{amsmath}
\usepackage{amssymb}
\usepackage{epic}
\usepackage{epstopdf}
\usepackage{graphicx}
\usepackage{hyperref}
\usepackage{multirow}
\usepackage{natbib}
\usepackage{overpic}
\usepackage{color}

\AtBeginShipout{%
  \ifnum\value{page}>1 %
    \typeout{* Additional boxing of page `\thepage'}%
    \setbox\AtBeginShipoutBox=\hbox{\copy\AtBeginShipoutBox}%
  \fi
}

\title[RA\MakeLowercase{i}SE III: 3C radio AGN energetics and composition]{RA\MakeLowercase{i}SE III: 3C radio AGN energetics and composition}

\author[R. J. Turner et al.]{
Ross J. Turner$^{1,2}$\thanks{Email: turner.rj@icloud.com},
Stanislav S. Shabala$^{1}$ and Martin G. H. Krause$^{1,3}$\\
$^{1}$School of Physical Sciences, University of Tasmania, Private Bag 37, Hobart, 7001, Australia\\
$^{2}$CSIRO Astronomy and Space Science, Post Office Box 76, Epping, New South Wales 1710, Australia\\
$^{3}$Centre for Astrophysics Research, School of Physics, Astronomy and Mathematics, University of Hertfordshire, College Lane,\\ Hatfield AL10 9AB, UK}

\date{Accepted 2017 November 13. Received 2017 November 13; in original form 2017 July 21}

\pubyear{2017}

\begin{document}
\label{firstpage}
\pagerange{\pageref{firstpage}--\pageref{lastpage}}
\maketitle

\begin{abstract}

Kinetic jet power estimates based exclusively on observed monochromatic radio luminosities are highly uncertain due to confounding variables and a lack of knowledge about some aspects of the physics of active galactic nuclei (AGNs). We propose a new methodology to calculate the jet powers of the largest, most powerful radio sources based on combinations of their size, lobe luminosity and shape of their radio spectrum; this approach avoids the uncertainties encountered by previous relationships. {The outputs of our model are calibrated using hydrodynamical simulations and tested against independent X-ray inverse-Compton measurements.} The jet powers and lobe magnetic field strengths of radio sources are found to be {recovered} using solely the lobe luminosity and spectral curvature, enabling the intrinsic properties of unresolved high-redshift sources to be inferred. By contrast, the radio source ages cannot be estimated without knowledge of the lobe volumes. The monochromatic lobe luminosity alone is incapable of accurately estimating the jet power or source age without knowledge of the lobe magnetic field strength and size respectively. We find {that, on average,} the lobes of the 3C radio sources have magnetic field strengths approximately a factor three lower than the equipartition value, inconsistent with equal energy in the particles and the fields at the $5\sigma$ level. The particle content of 3C radio lobes is discussed in the context of complementary observations; we do not find evidence favouring an energetically-dominant proton population.

\end{abstract}


\section{INTRODUCTION}
\label{sec:INTRODUCTION}

Knowledge of the kinetic power of active galactic nucleus (AGN) jets is vital for quantifying the energetics of AGN feedback and its effect on the host galaxy formation and evolution \citep{Croton+2006, Cattaneo+2009, Shabala+2009}. This kinetic jet-mode of feedback is dominant at the present epoch, whilst the radiative quasar-mode was prevalent in the early universe \citep[$z \sim 2$-$3$;][]{Fabian+2012}. Jet-mode feedback can explain the missing star formation in the most massive galaxies and clusters \citep{Croton+2006}, with the most powerful radio sources capable of suppressing {(or triggering)} star formation in not just their host but also other cluster galaxies \citep{Rawlings+2004, Shabala+2011}. {The relationship between jet power and accretion rate additionally enables investigation of the black hole accretion disk} \citep[e.g. ADAF or thin disk;][]{Shakura+1973, Narayan+1995, Turner+2015} and the contribution of the black hole spin to the production of the jets \citep{Meier+2001, Daly+2009, Daly+2016}. 

The kinetic powers of AGN jets have previously been estimated using radio source dynamical models \citep{Machalski+2004, Krause+2005, Shabala+2008, Punsly+2011, Turner+2015}, hot X-ray gas cavity observations \citep{Rafferty+2006, Birzan+2008, Cavagnolo+2010}, lobe expansion speeds {using spectral fitting} \citep{Daly+2012}, strong shocks around the lobes \citep{Croston+2009}, and measurements of the properties of the jet termination shocks \citep[hotspots;][]{Godfrey+2013}. However, there is a lack of reliable and generally applicable methods to estimate jet power. Many authors have attempted to simply convert radio luminosity directly to the jet kinetic power \citep[e.g.][]{Willott+1999, Cavagnolo+2010}. 
{However,} the radio luminosity must be modelled to correctly account for: (1) the low frequency cut-off in the energy spectrum of the radiating synchrotron electrons, (2) departures from the minimum energy assumption, (3) the particle composition of the lobe plasma, and (4) the filling factor of the synchrotron emitting radio lobe. \citet{Willott+1999} {discuss these difficulties and} attempt to constrain these sources of error but argue that up to two orders of magnitude uncertainty remain in the jet power.
Jet power--luminosity relations are also plagued by confounding variables such as the age-dependent electron loss processes \citep[e.g. inverse Compton, adiabatic and synchrotron;][]{KDA+1997}, source size \citep{Shabala+2013} and host environment \citep{KDA+1997, Barthel+1996}. Further, observed jet power--luminosity relations are often merely the result of strong selection effects \citep{Godfrey+2016}.

\citet{Turner+2015} presented a model, \emph{Radio AGN in Semi-analytic Environments} (RAiSE), which traces the expansion of lobed radio-loud AGNs throughout their evolution. This model considers sources exhibiting either \citeauthor{FR+1974} (FR) type I or II morphologies, including the transition between these classes. These mock radio sources are expanded into simulated environments predicted by semi-analytic models, with the shape of the gas density profile based on observations of local clusters \citep{Vikhlinin+2006}. The \citet{Turner+2015} model follows the evolution of radio sources from an initial supersonic expansion phase, through to a `transonic' phase where growth along the transverse (to the jet) axis slows. This leads to an elongation of the radio lobe along the jet axis, and thus a higher axis ratio\footnote{Our axis ratio is defined as the length of both lobes divided by the source width, and is related to the aspect (or axial) ratio of authors including \citet{Willott+1999} and \citet{Mullin+2008} through $A = 2R_{\rm T}$.}. At later times, the radio lobe mixes with the surrounding environment due to Rayleigh-Taylor and Kelvin-Helmholtz instabilities pinching the lobe from the active nucleus \citep[c.f.][]{Krause+2005, HK+2013}. Their dynamical model can therefore predict the filling factor of the lobe, removing this uncertainty from the \citet{Willott+1999} factor.

The \citet{Turner+2015} lobed FR-I/II dynamical model is complemented by a synchrotron emission model  based upon that of \citet{KDA+1997} and detailed in our companion paper \citep{Turner+2016}. {These dynamical and emissivity models are validated observationally by successfully reproducing the spectral ageing map of 3C436 and surface brightness map of 3C31 \citep[Figures 7 and 11 of][]{Turner+2016}.} Our synchrotron luminosity model considers the losses due to adiabatic expansion of the lobe, synchrotron radiation and inverse Compton scattering of cosmic microwave background photons. However, unlike the \citet{KDA+1997} model which must be applied to self-similar radio sources expanding into a power-law density profile, this method can be applied to a radio source in an arbitrary environment \citep[c.f.][]{Turner+2015}. Both models depend on the lobe magnetic field strength, lobe plasma particle composition, and the low-frequency cut-off to the electron energy spectrum. These parameters comprising the \citet{Willott+1999} uncertainty factor must still be constrained to enable a reliable conversion from the observed luminosity to the intrinsic jet power. 

In this paper, we develop a new technique for calculating the jet powers and ages of radio AGNs which does not require lobe length measurements but has the ability to constrain all the uncertainties in the luminosity calculation. Specifically, we can fit for departures from the lobe minimum energy assumption, the particle composition of lobe plasma (e.g. electrons, positrons and/or protons) and model the lobe filling factor, whilst observations constrain the low-frequency cut-off in the synchrotron electron energy spectrum.
Such a method enables the jet powers of unresolved sources at high-redshift to be accurately estimated. 
Our new jet power estimation method based on the spectral curvature of the radio lobe emission is detailed in Section \ref{sec:SPECTRAL EVOLUTION MODELS}. This method is applied to a sample of FR-II sources (described in Section \ref{sec:RADIO AGN DATA}) with the typical environments assumed for these sources detailed in Section \ref{sec:SIMULATED AGN HOST ENVIRONMENTS}. Our dynamical and synchrotron luminosity models are then calibrated based on observations in Section \ref{sec:CALIBRATION OF RADIO SOURCE MODEL}. Finally, in Section \ref{sec:PARAMETER ESTIMATION} we describe the various fitting algorithms used to determine the jet power, source age and lobe magnetic field strength based on differing observational constraints. 

The $\Lambda \rm CDM$ concordance cosmology with $\Omega_{\rm M} = 0.3$, $\Omega_\Lambda = 0.7$ and $H_0 = 70 \rm\,km \,s^{-1} \,Mpc^{-1}$ \citep{Komatsu+2011}, and spectral index, $\alpha$, defined by $S = \nu^{-\alpha}$ (for flux density $S$ and frequency $\nu$) {are} assumed throughout the paper.

\section{SPECTRAL EVOLUTION MODELS}
\label{sec:SPECTRAL EVOLUTION MODELS}

The jet power and age of radio AGNs in \citet{Turner+2015} were calculated using a maximum likelihood method based on the observed source size and $1.4 \rm\, GHz$ luminosity. In that work, the authors implicitly assumed the scaling between the jet power and luminosity through the choice of the low-energy cut-off to the synchrotron-emitting electrons in the lobes, and by equating the energy density in the particles and the magnetic field (i.e. equipartition). However, in this paper we attempt to derive jet powers using solely the physics incorporated in our dynamical and synchrotron luminosity models, in conjunction with known properties of observed radio AGNs. This means that we are no longer able to constrain the jet power using solely the source size and monochromatic luminosity, and must therefore find an additional observable capable of constraining the intrinsic AGN properties. In this paper, we use the shape of the observed radio spectrum as the additional constraint. The full treatment of the synchrotron emission derived in \citet{Turner+2016} is compared with simpler spectral ageing models, which enable the radio source spectrum to be readily parametrised in terms of a single variable. 

{The spectral ageing technique is not universally accepted as a reliable means of estimating the source dynamical age. The spectral age estimates for many powerful radio sources have been observed to be an order of magnitude lower than their dynamical ages \citep[e.g.][]{Rudnick+1994, Blundell+2000, Harwood+2015}. This discrepancy has been explained by a departure from the assumed equipartition field strength or the presence of non-radiating particles arising from the entrainment of thermal material. Recently, \citet{Turner+2016} showed that the spectral ageing models applied to resolved radio sources measure the age of the youngest electrons in a given region of the lobe; in the well-mixed lobes of FR-IIs the age of these youngest electrons is always at least a factor of a few less than the source dynamical age.} 
In Section \ref{sec:PARAMETER ESTIMATION}, we discuss the theoretical advantage of using the synchrotron spectrum to yield reliable jet power estimates.

\subsection{Spectral ageing models}
\label{sec:Spectral ageing models}

The spectral age of radio AGNs can be derived from the steepening in the observed synchrotron emission spectrum due to the synchrotron and inverse Compton loss processes \citep[e.g.][]{Jaffe+1973, Myers+1985, Alexander+1987, Murgia+1999, Jamrozy+2008}. The observed spectra are typically fitted using either the Jaffe-Perola \citep[JP;][]{Jaffe+1973} or Kardashev-Pacholczyk \citep[KP;][]{Kardashev+1962, Pacholczyk+1970} models to find the injection spectral index $\alpha_{\rm inj}$ and the break frequency $\nu_{\rm b}$, with the latter used to derive the spectral age. These models assume a single particle injection event produces the initial power-law distribution of electron energies $N(E) = N_0 E^{-s}$, where $N_0$ is a constant and $s = 2\alpha_{\rm inj} + 1$. The JP model assumes the pitch angle of electrons is isotropic only on short time-scales relative to the radiative lifetime, and that the electron density and magnetic field are homogeneous along the source depth. The KP model instead assumes all electrons maintain the same pitch angle during their radiative lifetimes; {however this model is unrealistic as any irregularities in the magnetic field will scatter electrons efficiently.}

{The other `standard'} model assumes the continuous injection of fresh particles over the lifetime of the source \citep[CI model;][]{Kardashev+1962, Pacholczyk+1970}, yielding a different distribution of electron energies. The high-frequency steepening of the CI and JP models can be thought of as limiting cases \citep{Carilli+1991}. The exponential cut-off of the JP model is the fastest the spectrum can steepen due to synchrotron losses whilst the CI model assumes minimum losses for a flatter spectrum. 

These standard spectral ageing models assume a uniform magnetic field strength $B_0$. \citet{Tribble+1991} proposed an analytical method to describe the synchrotron emission in a {locally} non-homogeneous magnetic field. This model assumes the field is a Gaussian random field as would be found if the magnetic field were the result of homogeneous, isotropic turbulence. The magnetic field strength distribution in this paradigm is that of a Maxwell-Boltzmann distribution \citep{Hardcastle+2013}. The standard models can be modified to assume this more realistic magnetic field distribution, and are referred to as the `Tribble' forms of the spectral ageing models \citep[i.e. TJP, TKP and TCI; e.g.][]{Harwood+2013}. 

The radio source spectrum can be well fitted by spectral ageing models, either for a lobe section with constant age electrons (e.g. JP and KP models) or for the entire radio lobe (e.g. CI model). These models include losses due to synchrotron radiation and the inverse-Compton scattering of cosmic microwave background photons, but compared to full synchrotron emission models \citep[e.g.][]{Turner+2016} they ignore losses due to the adiabatic expansion of the lobe. {The adiabatic losses can be neglected when fitting the spectral shape since they only translate the spectrum in the $\log\nu$--$\log S$ plane}.


\subsection{Synchrotron emissivity}
\label{sec:Spectral model derivation}

Following the method of \citet{Hardcastle+2013} and \citet{Harwood+2013} we derive expressions for the standard and Tribble form of the Jaffe-Perola, Kardashev-Pacholczyk and continuous injection models.
{The emissivity of the entire radio source electron population can then be found by integrating the single-electron emissivity \citep[see e.g.][]{Longair+2010} over the magnetic field, electron energies and pitch angles \citep[Equation 4 of][]{Hardcastle+2013},}

\begin{equation}
\begin{split}
J(\nu) = \int_0^\infty \! \int_0^\pi \! \int_0^\infty \frac{\sqrt{3} B e^3 \sin \xi}{8 \pi^2 \epsilon_0 c m_{\rm e}} F(x) N(E) p_\xi p_{\rm B} \,dE \,d\xi \,dB ,
\end{split}
\label{intensity}
\end{equation}

{where $p_\xi$ and $p_{\rm B}$ are the probability distributions of the pitch angle $\xi$ and the magnetic field strength $B$ respectively, $N(E)$ is the electron energy distribution, $e$ is the electron charge, $m_{\rm e}$ is the electron mass, $c$ is the speed of light, and $\epsilon_0$ is the permittivity of free space. The standard assumption for the pitch angle distribution of the electron population is that of isotropy, i.e. $p_\xi = \tfrac{1}{2} \sin \xi$. The magnetic field is generally assumed to be constant or a Maxwell-Boltzmann distribution as proposed by \citet{Tribble+1991}.
Finally, the single-electron synchrotron radiation spectrum $F(x)$ is defined as \citep{Rybicki+1979}}

\begin{equation}
F(x) = x \int_x^\infty K_{5/3}(y) \,dy ,
\end{equation}

{which is vanishingly small for $x \gtrsim 20$ and asymptotes to $F(x) = 2.15x^{1/3}$ for small $x$ \citep{Pacholczyk+1970}. Here $K_{5/3}$ is the modified Bessel function of order $5/3$, and $x$ is a dimensionless function of the frequency, field strength and energy:}

\begin{equation}
x = \frac{\nu}{\nu_{\rm b}} = \frac{4 \pi {m_{\rm e}}^3 c^4 \nu}{3 e E^2 B \sin \xi} ,
\end{equation}

{where $E$ is the electron energy, and $\nu_{\rm b}$ is the spectral break frequency beyond which (higher frequencies) the spectrum steepens due to the loss mechanisms.}

\subsubsection{Synchrotron age of electron population}

The synchrotron age of the radio source electron population is determined from the lobe magnetic field strength $B$ and the spectral break frequency $\nu_{\rm b}$ through the relation 

\begin{equation}
\tau = \frac{\upsilon B^{1/2}}{B^2 + {B_{\rm ic}}^2} \left[\nu_{\rm b} (1 + z) \right]^{-1/2} ,
\label{spectralage}
\end{equation}

where $z$ is the source redshift and $B_{\rm ic} = 0.318 (1 + z)^2 \rm\, nT$ is the magnitude of the magnetic field equivalent to the microwave background. Here, the constant of proportionality $\upsilon$ {for the JP model} is given by

\begin{equation}
\upsilon = \left(\frac{243\pi {m_{\rm e}}^5 c^2}{4 {\mu_0}^2 e^7} \right)^{1/2} ,
\end{equation}

in which $\mu_0$ is the magnetic permeability of free space. {This constant is reduced by a factor of 2.25 for the KP model \citep[e.g.][]{Nagai+2006}}.

\subsubsection{Jaffe-Perola and Kardashev-Pacholczyk models}

The Jaffe-Perola and Kardashev-Pacholczyk models generally assume electrons are impulsively injected at the hotspot. The injected packet of constant age electrons then propagates away from the hotspot; this leads to an age gradient across the lobe through multiple such injection events.
An initial power-law distribution of electron energies $N(E) = N_0 E^{-s} \delta(\tau)$, in the presence of synchrotron and inverse Compton losses, changes to \citep{Pacholczyk+1970, Longair+2010}

\begin{equation}
N(E) = 
\begin{cases}
N_0 E^{-s} (1 - \kappa E \tau)^{s - 2} &E < 1/\kappa \tau \\
0 &E \geqslant 1/\kappa \tau
\end{cases} ,
\end{equation}

where $\kappa$ is a constant such that $dE/d\tau = -\kappa E^2$ which takes account of both radiative and inverse-Compton losses. In this constant, we assume either the occurrence of synchrotron electron pitch angle scattering (JP model) or its absence (KP model). That is,

\begin{equation}
\kappa = \frac{2 \sigma_{\rm T}}{3 m_{\rm e}^2 \mu_0 c^3}
\begin{cases}
B^2 + {B_{\rm ic}}^2, & \text{for JP model} \\
B^2 \sin^2 \xi + {B_{\rm ic}}^2, & \text{for KP model}
\end{cases} ,
\label{ETKP}
\end{equation}

in which $\sigma_{\rm T}$ is the Thomson electron scattering cross-section. The radio spectra produced by these two spectral ageing models are shown in Figure \ref{fig:spectralmodels} assuming either a constant or Maxwell-Boltzmann magnetic field distribution.

\subsubsection{Continuous injection models}

Determination of the spectral age from the integrated lobe luminosity, such as in unresolved sources, necessitates the consideration of a continuous injection of electrons rather than just a single impulsive burst. {We note that by fitting the integrated spectra, continuous injection models are unable to describe inhomogeneities in the underlying physics; however, these models should provide a useful tool for understanding the integrated emission from the approximately homogeneous lobes of powerful FR-IIs.} Thus assuming a continuous injection of electrons, the initial power-law distribution of energies $N(E) = N_0 E^{-s}$ leads to the following electron energy distribution \citep{Longair+2010}:

\begin{equation}
N(E, \tau) = \frac{N_0 E^{-(s + 1)}}{\kappa (s - 1)}
\begin{cases}
1 - (1 - \kappa E \tau)^{s - 1} &E < 1/\kappa \tau \\
1 &E \geqslant 1/\kappa \tau
\end{cases} .
\label{NECI}
\end{equation}

Models describing a continuous injection of electrons can either assume that pitch angle scattering of the synchrotron electrons occurs (as for the JP model) or that it does not (as for the KP model). This assumption has negligible effect on the shape of the CI model spectrum for typical radio lobe properties; we therefore present a single continuous injection model allowing pitch angle scattering. {This model is applied to the spectra of a 3C sub-sample comprising 71 FR-II radio sources using 0.01-$10\rm\, GHz$ observations from \citet[][described in Section \ref{sec:RADIO AGN DATA}]{Laing+1980}; we find the CI model fits the spectra of 86\% of these objects at the 2$\sigma$ confidence level increasing to 93\% if a single outlying low-frequency point is excluded from the spectra of selected sources. The fit statistics for each object are included in Tables \ref{tab:fittedvalues} and \ref{tab:fittedvalues2} in the appendix. The lower frequencies in the remaining sources are generally noisy or have reduced flux densities due to free-free or synchrotron self-absorption. These are well fitted upon supplementing the \citet{Laing+1980} observations with additional literature data and excluding the low-frequency spectral curvature from the fit.}

{\citet{Harwood+2017} similarly investigated the ability of the CI model to fit the spectra of a range of radio sources and compared the resulting spectral ages to those derived using the JP model in their BRATS software package. The CI model fits the high-frequency end of the spectra well for the four (active) FR-IIs in their sample, but cannot fit the lower frequency observations of two of these objects, consistent with the results of our fitting for the same data). 
\citet{Harwood+2017} also found the spectral ages measured using the CI model differ by up to a factor of six compared to those calculated for resolved sources using the JP model. This inconsistency is not an indication of problems with the CI model age estimates; rather \citet{Turner+2016} found that the JP model yields spectral ages a factor of a few different from the dynamical age in well-mixed lobes of FR-IIs. This discrepancy occurs due to a violation of the JP model assumption that the spectrum results from a single injection of electrons, despite the oldest regions of the lobe comprising electron populations differing by a factor of up to a few in age.}

\subsection{Independence of magnetic field}
\label{sec:Independence of magnetic field}

\begin{figure}
\begin{center}
\includegraphics[width=0.48\textwidth]{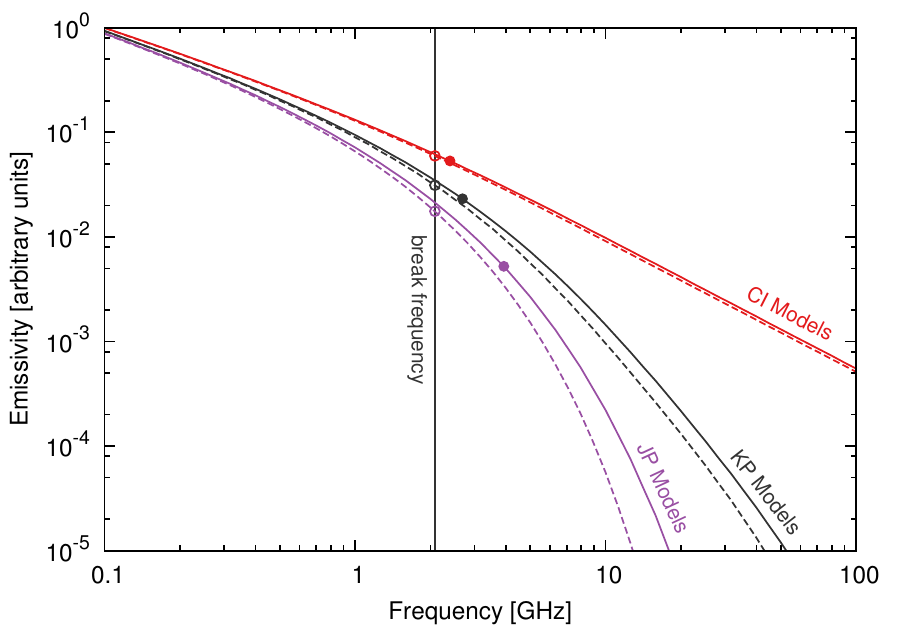} 
\end{center}
\caption{Spectral shape expected for the Jaffe-Perola (JP--purple), Kardashev-Pacholczyk (KP--grey) and Continuous Injection (CI--red) models, assuming either a constant (dashed) or Maxwell-Boltzmann distribution (Tribble model--solid) for the magnetic field. These models assume a field strength of $B = 10\,\rm nT$, an injection index of $s = 2.5$, and correspond to a spectral age of $\tau = 30\,\rm Myrs$. The vertical line marks the $\sim 2\,\rm GHz$ break frequency of the six spectral models (KP break scaled to match JP and CI breaks on plot); the circles show the break frequency derived by fitting the standard constant magnetic field forms of the models to the spectra over the 0.1 to $100\rm\, GHz$ frequency range.}
\label{fig:spectralmodels}
\end{figure}

{The structure and magnitude of the magnetic field can potentially introduce a large source of error into the spectral ageing technique.
In this section, we quantify the effect the structure of the lobe magnetic field has on the shape of the JP, KP and CI model spectra. These models are plotted in Figure \ref{fig:spectralmodels} for two magnetic field distributions: (1) the constant field assumption of the standard form of the models, and (2) a Maxwell-Boltzmann distribution (Tribble form). Local inhomogeneities in the magnetic field lead to a spectrum arising from populations of electrons with spectra described by the standard form of the ageing models, but whose break frequencies are shifted due to the change in field strength. The spectra for the Tribble ageing models thus have more high-energy electrons and as a result have their spectral breaks shifted to higher frequencies. However, the continuous injection model spectra are largely unaffected by inhomogeneities in the magnetic field because a significant population of high-energy electrons already contribute to the spectra. That is, local inhomogeneities in the magnetic field have negligible effect on the shape of the integrated radio spectrum, in stark contrast to the large difference seen in the JP spectra investigated by \citet{Hardcastle+2013}.}

{The difference in the standard and Tribble-CI spectra is quantified by fitting the standard form of the model to a Tribble-CI spectrum over a typical frequency range of 0.1 to $100\,\rm GHz$; we find the break frequency fitted in this scenario is a factor of 1.15 greater than for the original Tribble spectrum. This systematic error leads to an underestimation of magnetic field strengths and thus equipartition factors by a factor of 1.05. Local inhomogeneities in the magnetic field are therefore not expected to appreciably alter the CI spectra or any results derived from the fitted spectral index and break frequency.}

The standard form of the continuous injection model, and the JP and KP models, assume a constant lobe magnetic field strength. The integral in Equation \ref{intensity} can therefore be reduced from a triple to a double integral, with the field strength as a scaling factor. However, the single-electron synchrotron radiation spectrum $F(x)$ is dependent on the strength of the magnetic field. We seek to remove this dependence by parametrising the integral in terms of $x$, leaving only frequency independent terms involving the field strength. The integrated emissivity of the continuous injection model can thus be written in terms of integrals over only $x$ and the pitch angle. That is,

\begin{equation}
J(\nu) = J_0 \nu^{-s/2} \int_0^{\pi/2} \sin^{(s + 4)/2} \xi \int_0^\infty F(x) \mathcal{N}(x) dx d\xi ,
\end{equation}

where $J_0$ is a frequency-independent constant, and $\mathcal{N}(x)$ is the frequency-dependent component of the electron energy distribution. For the continuous injection model this component of the electron energy distribution is then given by

\begin{equation}
\mathcal{N}(x) = x^{(s - 2)/2}
\begin{cases}
1 - x^{(1 - s)/2}(x^{1/2} - \iota^{1/2})^{s - 1} &x > \iota \\
1 &x \leqslant \iota
\end{cases} ,
\end{equation}

where $\iota(\nu, \xi) = \nu/(\nu_{\rm b} \sin \xi)$ is the value of $x$ corresponding to an energy of $E = 1/\kappa \tau$ (cf. Equation \ref{NECI}). Similar expressions can be obtained for the standard JP and KP models \citep{Nagai+2006}. 

The magnetic field strength of radio sources therefore does not need to be quantified in order to observationally constrain the injection index $s$ of the electron ensemble or the spectral break frequency $\nu_{\rm b}$. 
However, the synchrotron age of the electron population does depend on the field strength through Equation \ref{spectralage}. 

Different techniques for estimating the lobe magnetic field strength exist in the literature. 
The most basic approach is to determine the field strength from the level of synchrotron emission under the assumption of equal energy density in the particles and the field \citep{Burbidge+1956}. However, in low-powered FR-Is which are expected to be in pressure equilibrium with their surroundings, this approach predicts underpressured lobes \citep{Croston+2008} suggesting either the field may not be in equipartition {or that additional pressure is provided by entrained and heated thermal plasma}.
A more direct approach uses the emission from relativistic electrons inverse-Compton scattered to X-ray energies to calculate the energy density in the synchrotron electrons. The magnetic field strength is derived by comparing the observed X-ray emissivity to model predictions derived from the level of synchrotron emission. \citet{Hardcastle+2002}, \citet{Croston+2004, Croston+2005} and \citet{Ineson+2017} all find sub-equipartition magnetic field strengths in the lobes of FR-IIs. Spectral ages calculated assuming fields in equipartition will thus generally be greatly underestimated. In this work we instead use lobed FR-I/II dynamical model of \citet{Turner+2015} to fit the magnetic field strength based on the source size and break frequency, additionally yielding estimates for the age and jet power.

\subsection{Dynamical versus synchrotron ages}
\label{sec:Dynamical versus synchrotron ages}

{The ability of the CI model to provide a good empirical fit to observed spectra (and hence quantify their curvature) is tested using simulated spectra for radio sources with known field strengths and particle content}. This enables the spectral age to be fitted and compared to modelled sources with a known dynamical age. The spectra of these {simulated} radio sources are generated using the lobed FR-I/II dynamical model of \citet{Turner+2015} and synchrotron emissivity model of \citet{Turner+2016}, which includes all {radiative} loss mechanisms and the temporal evolution of the lobe magnetic field strength. The \citet{Turner+2016} model provides a more complete description of the radio spectrum than standard spectral ageing models, however the CI spectrum is preferable for parameter fitting, when a good approximation, for two reasons: (1) the computation time of the \citet{Turner+2016} model increases with the number of observed frequencies, and (2) does not provide a single measure of the curvature (e.g. $\nu_{\rm b}$) increasing the complexity of any parameter estimation method.

\subsubsection{RAiSE II integrated luminosity model}

The formalism for the integrated radio lobe luminosity model \citep[Section 2 of][]{Turner+2016}, which is applied to the lobed FR-I/II dynamical model, is based on the work of \citet{KDA+1997}. This model calculates the synchrotron emissivity based on the magnetic field strength at the time of electron injection rather than assuming a population average as for the approximating continuous injection model discussed in Section \ref{sec:Independence of magnetic field}. 
The general form for the lobe luminosity in the \citet{KA+1997} synchrotron emission model is given by

\begin{equation}
\begin{split}
L(\nu, s, t) = K(s) \nu^{(1 - s)/2} &\frac{q^{(s + 1)/4}}{(q + 1)^{(s + 5)/4}(w + 1)} \\
&\quad\quad p(t)^{(s + 5)/4} V(t) \mathcal{Y}(\nu, s, t) ,
\end{split}
\label{luminosityloss}
\end{equation}

where $q = u_{\rm B}/u_{\rm e}$ is the ratio of the energy density in the magnetic field to that in the synchrotron-emitting leptons, $w = u_{\rm t}/u_{\rm e}$ is the ratio of the energy density in the {non-radiating} particles and the {leptons}, $p(t)$ is the lobe pressure, $V(t)$ its volume, $\mathcal{Y}(\nu, s, t)$ the loss function, and $K(s)$ a radio source constant. This source-specific constant is a function solely of physical constants, the injection spectral index, the adiabatic index of the lobe $\Gamma_{\rm c}$, and the minimum Lorentz factor $\gamma_{\rm min}$ of the injected synchrotron electron population. In this work, we assume lobes comprising a relativistic pair-plasma potentially supplemented by a non-relativistic thermal fluid (i.e. $4/3 \leqslant \Gamma_{\rm c} \leqslant 5/3$), though we find no more than a $0.1 \rm\, dex$ variation in the fitted parameters between the extremes of entirely relativistic or non-relativistic fluids. Finally, the low-energy cut-off to the electron population is observationally constrained (see Section \ref{sec:CALIBRATION OF RADIO SOURCE MODEL}). 

The loss function is the fractional reduction in luminosity relative to the lossless case due to the lobe adiabatic expansion, synchrotron radiation, and the inverse Compton scattering of cosmic microwave background photons. \citet{Turner+2016} model the loss function both as a function of position in the lobe (c.f. JP and KP models) and for the integrated luminosity (c.f. CI models). Integrated multi-frequency radio observations, such as for unresolved lobes, can therefore be simulated for radio sources evolved to any age using the lobed FR-I/II dynamical model.

\subsubsection{Simulated radio spectra}

\begin{figure}
\begin{center}
\includegraphics[width=0.48\textwidth]{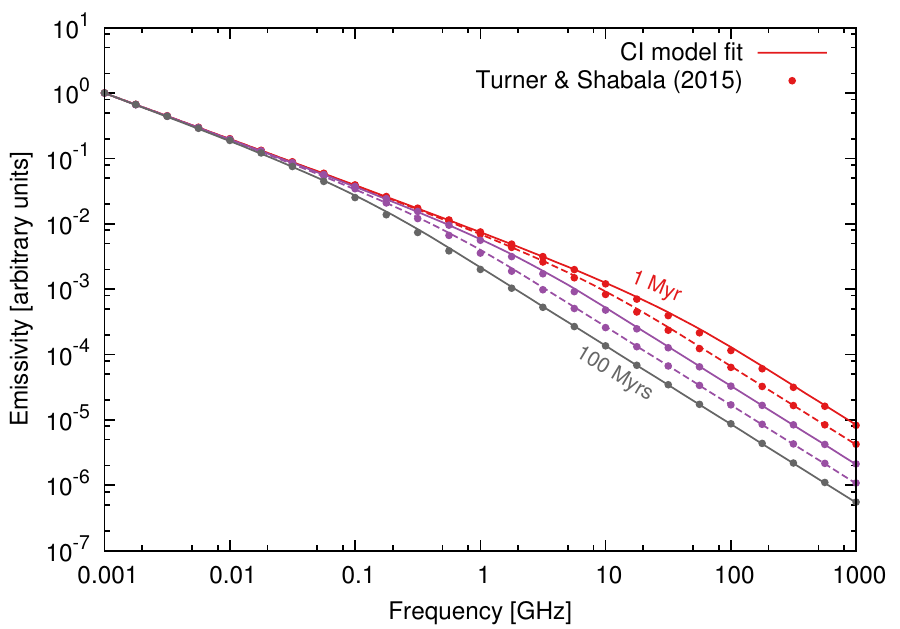} 
\end{center}
\caption{Fits of the standard continuous injection (CI) model to the simulated radio source spectra from our lobed FR-I/II dynamical model. These spectra are for a radio source with jet power of $10^{34} \rm\, W$ inhabiting a $3\times 10^{10} \rm\, M_\odot$ host galaxy, equipartition lobe field strength, and electron energy injection index of $s = 2.4$. The points are scaled luminosities calculated using our dynamical model at $0.25\rm\,dex$ steps in frequency, coloured by the dynamical age of the source. The dynamical age increases top-to-bottom from $1\rm\,Myr$ (red -- top) through to $100\rm\,Myrs$ (grey -- bottom) in log-space steps of $0.5\rm\,dex$. The solid and dashed lines both show the best fits obtained using the CI model for each dynamical age.}
\label{fig:spectralfits}
\end{figure}

The radio spectra of a wide range of radio sources are simulated using our combined lobed FR-I/II dynamical model and synchrotron luminosity model. Jet powers of between $10^{34}$ and $10^{40}\rm\, W$ are simulated for radio sources inhabiting $3\times10^{10}$ to $10^{12}\rm\,M_\odot$ stellar mass hosts, whose lobes have initial axis ratios between $A = 2$ and $8$, and with magnetic field strengths ranging from equipartition to a factor of 100 lower. The evolutionary tracks of these sources are analysed from $1$ to $1000\rm\, Myrs$, with simulated spectra produced at regular time-steps. The simulated spectra of a typical source are shown in Figure \ref{fig:spectralfits} for five time-steps, with the best continuous injection model fits for each age overplotted. The standard spectral ageing models (i.e. non-Tribble forms of JP, KP and CI models) have two free parameters, the injection spectral index $\alpha_{\rm inj}$ and the break frequency $\nu_{\rm b}$, in addition to a constant term. The relevant spectral model is fitted to these multi-frequency radio lobe spectrum simulations using a weighted log-space least squares regression to determine the best estimates and uncertainties for these parameters.
The parametrised form of the continuous injection model fits our simulated spectra extremely well for all time-steps, suggesting any comparison between the fitted spectral and simulated dynamical ages will be well-founded. 

\begin{figure}
\begin{center}
\includegraphics[width=0.48\textwidth]{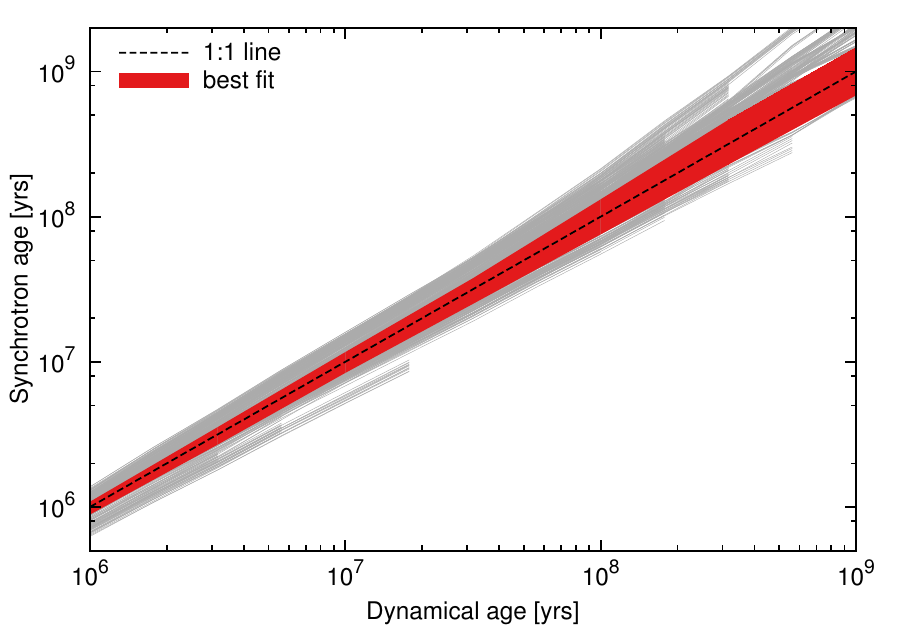} 
\end{center}
\caption{Synchrotron age as a function of the dynamical age for a broad range of simulated radio galaxies. The radio source jet power, magnetic field strength, synchrotron energy injection index, and host galaxy properties are varied in these simulations (grey lines). The thick red line plots the $1\sigma$ uncertainties in the synchrotron-dynamical age relationship.}
\label{fig:dynvssynage}
\end{figure}

The best-fitting models to the simulated spectra yield estimates of the spectral break frequency and injection spectral index. The former is combined with the magnetic field strength to determine the age of the synchrotron electron population via Equation \ref{spectralage}. For these simulated radio sources the magnetic field strength is found directly from the lobe pressure and equipartition factor computed as part of the dynamical model. The evolutionary tracks of all the simulated radio sources are plotted in Figure \ref{fig:dynvssynage} in synchrotron--dynamical age space. The fitted synchrotron electron age is consistent with the dynamical age of the simulated radio source for all ages within the $1\sigma$ uncertainties. We are therefore justified in using the continuous injection model to fit the curvature of observed spectra based on the single parameter break frequency statistic, and further to equate the spectral age obtained using Equation \ref{spectralage} with the dynamical age.

The strong relationship found between the synchrotron and dynamical ages is contingent on accurate measurements of the lobe magnetic field strength. We note that mismeasurement of the magnetic field strength in existing spectral ageing methods leads to inaccuracies in the age as $\tau \propto B^{-3/2}$.
Current measurement techniques require the lobe-related X-ray emission be measured, necessitating the surrounding X-ray emitting hot gas must be excluded \citep[e.g.][]{Hardcastle+2002, Croston+2004, Croston+2005}.
However, compact sources and much of the high-redshift population are unresolved making this approach impractical at present. In this work, the magnetic field strength will be fitted using a Bayesian approach, constrained using other observables including the source size and radio spectrum of the source.

\section{POWERFUL RADIO AGN SAMPLE}
\label{sec:RADIO AGN DATA}

In this section, we apply the combined dynamical and synchrotron luminosity model to a well-studied sample of powerful radio galaxies.
\citet{Mullin+2008} presented a catalogue\footnote{http://zl1.extragalactic.info/} of 98 $z < 1$ \emph{Third Cambridge Catalogue of Radio Sources} (3C). Each object includes the measured 178\,MHz radio lobe luminosity, $R$-band optical magnitude (for approximately three quarters of the sample), and the largest linear source size, width and axis ratio of each lobe. The \citet{Mullin+2008} dataset is compiled from the complete flux-limited sample of \citet{Laing+1983} which includes all 3C sources with total $178\rm\,MHz$ flux densities $S_{178} > 10.9\rm\,Jy$. {These 3C galaxies are further cross-matched with $K$-band near-infrared magnitudes and multi-frequency radio observations to enable estimation of the host stellar mass and fitting of the spectral curvature respectively}. These spectral fits will provide an alternative method of constraining the source age, in contrast to the sole use of the size parameter \citep{Turner+2015}.

\subsection{Spectral break frequencies}
\label{sec:Spectral break frequency fitting}

\begin{figure}
\begin{center}
\includegraphics[width=0.46\textwidth]{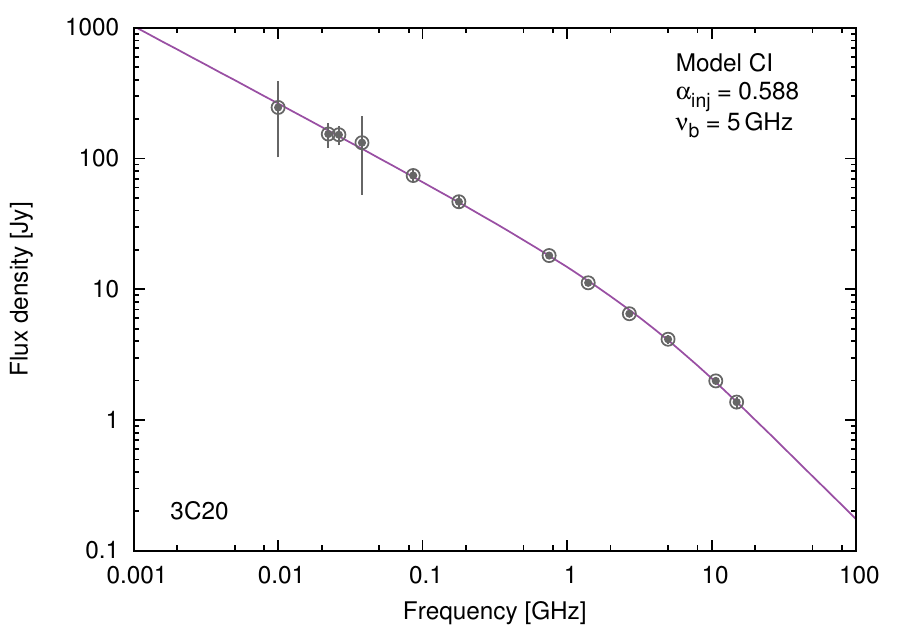} 
\end{center}
\caption{Spectral fit to the lobe synchrotron emission of the typical radio source 3C20. The observed flux densities taken from \citet{Laing+1980} are plotted with grey circles and $2\sigma$ errorbars. The fit to these multi-frequency observations assumes the parametrised form of the continuous injection model (Section \ref{sec:Independence of magnetic field}), with the fitted injection spectral index $\alpha_{\rm inj}$ and break frequency $\nu_{\rm b}$ parameters stated on the plot.}
\label{fig:KPJPfit}
\end{figure}

The {integrated} radio spectra of 88 3C sources in the full \citet{Mullin+2008} sample are fitted using the parametrised form of the continuous injection model, with multi-frequency data taken from \citet{Laing+1980}. 
{The limited frequency range of their observations makes it inevitable that only sources with a break of between 0.1-$10\rm\, GHz$ will be fitted; it is vital that spectra with no clear curvature in this range are excluded. We use additional archival observations to confirm the fitted curvature is real and not due to random noise in a few points. The break frequency is confidently fitted for 37 sources and a lower limit of $\sim 10\rm\, GHz$ is claimed for a further 34 objects; the injection index is fitted for all 71 of these objects. The spectral fit obtained for a typical radio source with a break frequency between 0.1 and $10\rm\, GHz$ is plotted in Figure \ref{fig:KPJPfit}. 
This limited break frequency range will bias any sample to the oldest radio sources at low-redshift, and younger objects at higher redshifts; however, a complete sample can be constructed by applying informed prior probability distributions (from the biased sample) to remaining objects with break frequency lower limits.} 

{The spectra of some of the sources with multi-frequency observations are not fitted for one of the following reasons: (1) a sharp low-frequency turnover is present, due to either synchrotron self-absorption \citep{Kellermann+1966} or free-free absorption (\citealt{Bicknell+1997}; 3C268.3, 3C295), (2) the radio source has irregular lobe morphology (3C215, 3C351, 3C433), (3) the lobe has no or incomplete axis ratio measurements (3C254, 3C299, 3C321), or (4) the high-frequency emission is dominated by the core or Doppler-boosted components associated with the jets (3C109, 3C207, 3C216, 3C220.1, 3C263, 3C275.1, 3C334, 3C382, 3C390.3).} 
{These latter nine sources are removed based on the high-resolution 1.5-$8.5\rm\, GHz$ observations compiled by \citet{Mullin+2008}; the cores of these sources comprise more than $10\%$ of their total emission compared to a median core-to-lobe emissivity ratio of less than $1\%$ for the rest of the sample.} We note that it may be possible to correct the integrated emission by removing the core contribution based on literature observations at each observing frequency.

\subsection{Stellar masses}
\label{sec:Stellar masses}

The radio source evolution model developed by \citet{Turner+2015} requires some knowledge of the host environment. For massive AGN hosts ($M_\star > 3\times10^{10} \rm\, M_\odot$) the gas density was shown to be well approximated by a sequence of power law profiles whose scaling is based on the properties of the host galaxy halo. We discuss the applicability of this model for the environments of our 3C sample in Section \ref{sec:SIMULATED AGN HOST ENVIRONMENTS}. Empirical relations determined from semi-analytic galaxy evolution models \citep[SAGE;][]{Croton+2006, Croton+2016} are used to determine the scaling of these profiles from the observed host stellar mass and redshift. Stellar masses must therefore be determined for our sub-sample of 3C galaxies.

\begin{figure}
\begin{center}
\includegraphics[width=0.48\textwidth]{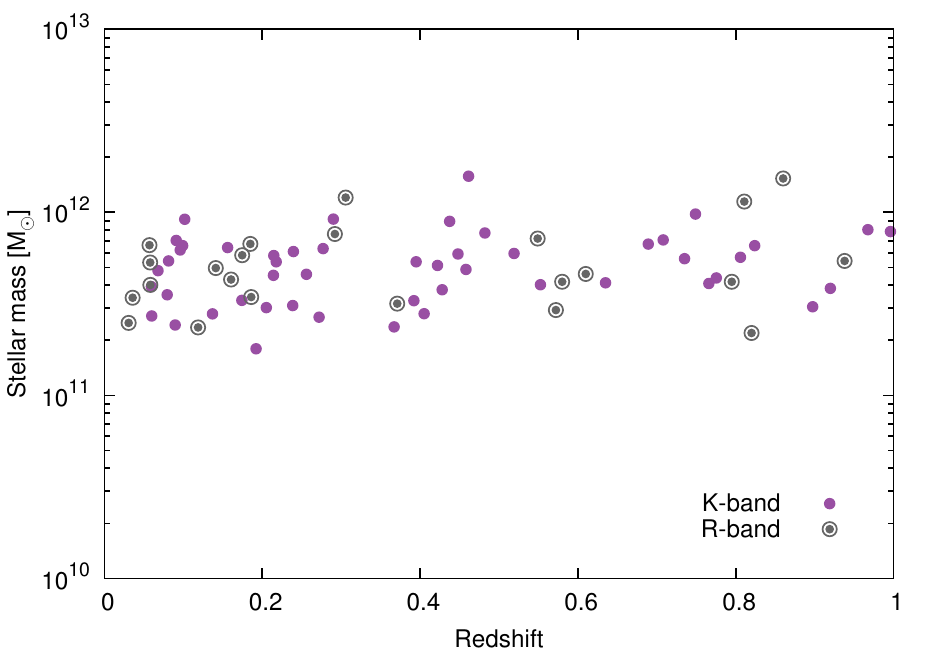} 
\end{center}
\caption{The stellar masses estimated for the 73 3C sources as a function of redshift. The 50 sources plotted with filled purple circles have their galaxy stellar masses calculated directly from observed $K$-band magnitudes \citep{McLure+2006}, whilst a further 23 galaxies have their masses estimated from $R$-band magnitudes (black circles).}
\label{fig:stellarredshift}
\end{figure}

{The stellar masses of our 3C sources are obtained using the observed $K$- and $R$-band magnitudes from \citet{Willott+2003}. \citet{McLure+2006} have previously estimated the stellar masses of the 3C sources with $K$-band magnitudes in the \citet{Mullin+2008} sample. The stellar masses for many of the approximately half remaining sources in the sample without $K$-band magnitudes are estimated using $R$-band magnitudes and the empirical relationship of \citet{Longhetti+2009}. This relationship is calibrated for $R$-band using the portion of the \citet{McLure+2006} sample with both $K$- and $R$-band magnitudes.} 
We thus have stellar mass estimates for 73 of the 98 3C radio sources examined by \citet[57 of the 71 with fitted spectra;][]{Mullin+2008}; these mass estimates are plotted as a function of redshift in Figure \ref{fig:stellarredshift}. These 73 3C sources have a median stellar mass of $M_\star = 4.9\times10^{11} \rm\, M_\odot$, and show only a weak dependence on redshift. Further, the stellar masses of these 3C sources are very similar at all redshifts with a standard deviation of only $0.19\rm\, dex$. Moreover, in Section \ref{sec:SIMULATED AGN HOST ENVIRONMENTS} we show that radio source expansion is only weakly affected by the host environment. A population average may therefore be sufficient in this work for the remaining 25 radio sources lacking X-ray or optical stellar mass estimates, {enabling these object to be included in the sample}.

\subsection{Viewing angle corrections}
\label{sec:Viewing angle corrections}

The observed linear sizes and axis ratios of radio sources will be affected by the viewing angle. At small angles the linear size and axis ratio should also be close to the intrinsic value ($\propto \cos d\psi$, {where $d\psi$ is the angle between the line-of-sight and the normal to the jet axis}). Moreover, since the typically rounded end of radio lobes comprise a spherical sector the same linear size is expected to be measured any viewing angle within $d\psi < 1/\mathcal{A} \rm\: radians$ of the normal to the jet axis, where $\mathcal{A}$ is the observed axis ratio. For a typical observed axis ratio of $\mathcal{A} = 6$ the linear size should be unchanged for a viewing angles within 15 degrees of the normal to the jet axis. {Combining these two effects, the length of a typical radio source viewed at 30 degrees from normal to the jet axis would be observed as $97\%$ of its intrinsic length, reducing to $87\%$ for a viewing angle of 45 degrees.}
Objects viewed at more extreme angles may need to have their lobe lengths corrected for the viewing angle. {The inclination angle is known to be anti-correlated with the fractional flux density in the core \citep{Orr+1982}; radio sources at extreme viewing angles should therefore have already been removed from the sample by our exclusion of objects with strong core emission.}

\section{SIMULATED AGN HOST ENVIRONMENTS}
\label{sec:SIMULATED AGN HOST ENVIRONMENTS}

Knowledge of the environments into which radio sources expand is critically important when modelling their expansion and evolution. The radio source environment may be determined not only by the host galaxy but also the properties of the surrounding cluster. {Observations of the hot X-ray gas surrounding 3C sources have been undertaken primarily at low-redshift \citep[e.g.][]{Allen+2001, Massaro+2012}. However, the radio sources examined in this work extend out to $z = 1$ (with a median of $z \sim 0.6$) necessitating the use of semi-analytic galaxy evolution models to describe the environments of many of these objects.} 
In this section, we describe a model for the gas density and temperature of the environments surrounding the radio sources in the 3C sub-sample.

\subsection{Sub-halo or large-scale cluster environment?}
\label{sec:Sub-halo or large-scale cluster environment}

Within the framework of our model, the environments surrounding AGNs in central brightest cluster galaxies (BCGs) can be modelled solely using the host stellar mass to infer the hot gas mass of the central dark matter halo. However, for AGNs in smaller non-central galaxies the contribution from both the central halo and sub-halo associated with the host galaxy may need to be considered. The modelled evolution of any non-central sources inhabiting rich cluster environments will be affected if the cluster density contribution exceeds that of the sub-halo, causing the gas density profile to flatten greatly. The source evolution in the flat\footnote{The cluster contribution to the density profile is affected largely by the orientation of the jets relative to the cluster centre. Lobes pointing away from the centre will experience a declining profile whilst those towards it a rising density. Hence, assuming the jet angle is random the typical cluster environment at the AGN is an initially flat, then very slowly declining gas density profile.} cluster density profile will be very different to that of the rapidly declining sub-halo profile. The host galaxy's location within the cluster may therefore greatly affect the AGN environment, not only the mass of the associated sub-halo. 

The mass of the cluster dark matter halo can be determined using thermal X-rays, weak gravitational lensing, or by combining the masses of the sub-haloes associated with every galaxy in the cluster \citep[e.g.][]{Yang+2007}, though in practice these methods are only possible at low redshifts.
The location of our AGN within their broader cluster environment is examined using the properties of approximately $100\,000$ simulated galaxies taken from the Semi-Analytic Galaxy Evolution \citep[SAGE;][]{Croton+2016} model. 
The most massive non-central galaxies simulated using SAGE are all predicted to have stellar masses less than $M_\star = 2.1\times10^{11} \rm\, M_\odot$, compared to the observed mean of $4.9\times10^{11} \rm\, M_\odot$ for our 3C sample. The probability of a given 3C radio source in our sample (based on the mass distribution) inhabiting a non-central galaxy is only $3.5\%$, with a vanishingly small probability of there being numerous non-BCG radio AGNs. {Moreover, these results are consistent with the analysis by \citet{Best+2000} of the environments surrounding high-redshift ($0.6 < z <1.8$) 3C sources.} The environments surrounding the radio sources in this work will therefore be modelled assuming the host is the central BCG.

\subsection{Halo host environment}
\label{sec:Sub-halo host environment}

The environment into which these 3C radio sources expand can be modelled using an assumed gas density profile for the central halo (for the assumed BCGs) scaled by its total gas mass. The adopted cluster density and temperature profiles are described in the following.

\subsubsection{Cluster gas density profile}

\begin{figure}
\begin{center}
\includegraphics[width=0.48\textwidth]{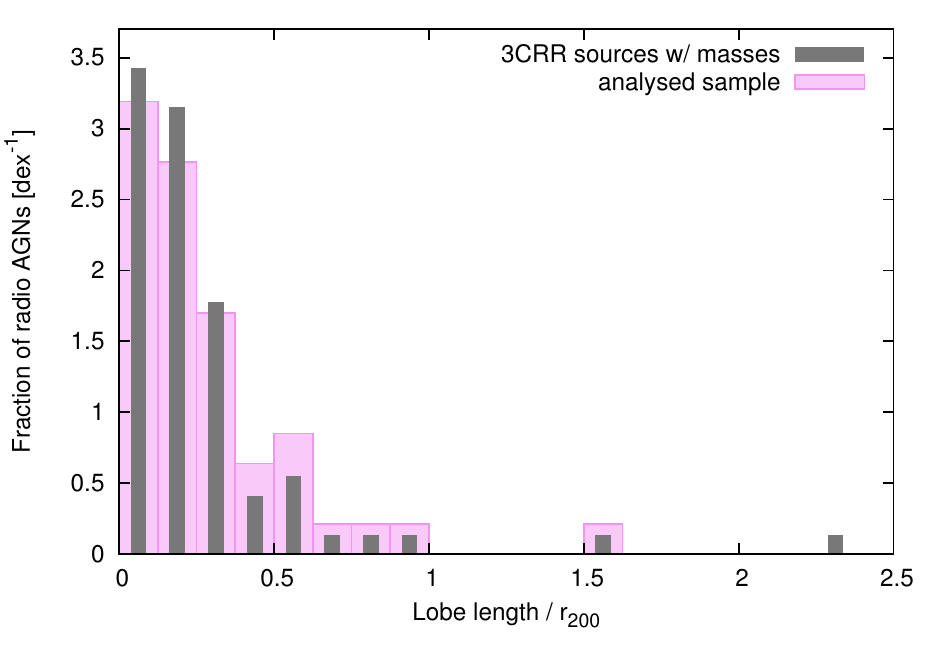} 
\end{center}
\caption{Distribution of the ratio of the observed lobe length (source length/2) and halo virial radius ($r_{200}$) for each 3C source in our sample. This distribution is shown both for all sources with mass estimates (grey) and only those included in the analysis in Section \ref{sec:PARAMETER ESTIMATION} (pink).}
\label{fig:drvir_hist}
\end{figure}

The central halo gas density profile is modelled using the realisations of the \citet{Vikhlinin+2006} profile as discussed by \citet{Turner+2015}. These density profiles are observed to have very similar shapes when scaled by the gas mass associated with the halo and its virial radius. These cluster density profiles are only fitted out to approximately the virial radius \citep[Figure 17 of][]{Vikhlinin+2006}. The size of our sources relative to the virial radius of their host halo is examined in Figure \ref{fig:drvir_hist}. We find that only two of the 3C sources in our sample expand beyond the virial radius, suggesting the fitted cluster profile is generally suitable throughout the radio source evolution. 

We derive relationships for the gas mass and virial radius as a function of the halo mass to simplify the parametrisation of the host environment. Further, a conversion from the observed stellar mass to this halo mass can be obtained using the mock galaxies of SAGE. 
These relationships enable the host environment of each radio source to be realistically modelled using solely the observed host galaxy stellar mass.
The mass of the AGN host galaxy's dark matter halo is related to its virial radius $r_{\rm vir}$ through

\begin{equation}
M_{\rm vir} = \frac{100}{G} H^2(z)\:\! {r_{\rm vir}}^3 ,
\end{equation}

where $G$ is Newton's gravitational constant, and $H(z)$ is the Hubble constant at redshift $z$. The baryonic fraction in the massive central haloes inhabited by 3C sources is observed to approach the universal value \citep{McGaugh+2010, Gonzalez+2013}. We therefore assume the gas mass of the halo environment is $M_{\rm gas} = f_{\rm b} M_{\rm vir}$ for cosmic baryon fraction $f_{\rm b} = 0.15$ \citep{Planck+2014}.

The average gas densities of these AGN host haloes are independent of the host mass, and are a function solely of redshift through the Hubble constant, i.e. $\bar{\rho}_{\rm gas} = 100 f_{\rm b} H^2(z)/G$. However, the environments into which the radio sources expand are by no means identical since the density scales with increasing virial radius and thus mass. For a rough approximation of the gas density profile of the form $\rho(r) = \rho(r_0) [r/r_0]^{-\beta}$, the gas density at some scale radius $r = r_0$ is related to the virial radius by

\begin{equation}
\rho(r_0) = (3 - \beta) \bar{\rho}_{\rm gas} \left(\frac{r_{\rm vir}}{r_0} \right)^\beta ,
\end{equation}

where the exponent $\beta$ is typically no greater than two \citep{KA+1997}. The gas density at this arbitrary radius reached by the expanding radio source increases with the halo mass as $\rho(r_0) \propto {M_{\rm vir}}^{\beta/3} \sim {M_\star}^{\beta/3}$. The gas density is therefore moderately insensitive to the observed stellar mass, varying by less than a factor of two within the $1\sigma$ level of uncertainty for our stellar mass estimates (Section \ref{sec:Stellar masses}). Further, the uncertainty in the baryon fraction in massive clusters is no greater than this factor \citep{Gonzalez+2013}. Propagating this $1\sigma$ error in the host gas density yields a $0.01\rm\,dex$ uncertainty in jet power estimates, $0.14\rm\,dex$ in source ages, and $0.08\rm\,dex$ in our equipartition factors. This insensitivity to the observed host stellar mass (and environment in general) enables a typical 3C environment to be assumed for the remaining 25 objects in our sample lacking stellar mass estimates.

\subsubsection{Host halo temperature}
\label{sec:Sub-halo temperature}

The temperature of the hot gas in clusters is known to scale with the halo mass as $T \propto {M_{\rm vir}}^{0.67}$ \citep[e.g.][]{Vikhlinin+2006}, in line with the virial temperature of the halo. \citet{Turner+2015} argued that the gas temperatures in smaller hosts may be higher due to heating from the AGN, however the 3C host galaxies examined here have stellar masses $0.5 \rm\, dex$ higher than those in that paper. Such massive AGN hosts are observed by \citet{OSullivan+2001} to have comparable gas temperatures to other clusters. In this work, we will therefore consider hot gas temperatures which scale with halo mass as $T \propto {M_{\rm vir}}^{0.67}$ and neglect the constant temperature assumption for all but the lowest mass hosts.

\section{CALIBRATION OF RADIO SOURCE MODEL}
\label{sec:CALIBRATION OF RADIO SOURCE MODEL}

The spectral ageing process discussed in Section \ref{sec:SPECTRAL EVOLUTION MODELS} requires knowledge of the lobe plasma characteristics and radio source dynamics to derive the physical properties of real active galaxies. In this section, we assess the validity of the \citet{Turner+2015} dynamical model by comparing predicted radio source evolution to the detailed, albeit more computationally intensive, hydrodynamical simulations of \citet{HK+2013}. The axis ratio evolution predicted by our the lobed FR-I/II dynamical model is compared to that of their simulations in addition to the measured 3C axis ratios. The \citet{Turner+2016} synchrotron emission model, which is based on the work of \citet{KDA+1997}, is also calibrated in this section using various radio source observations.

\subsection{Validation of radio source dynamics}

\begin{figure}
\begin{center}
\includegraphics[width=0.48\textwidth]{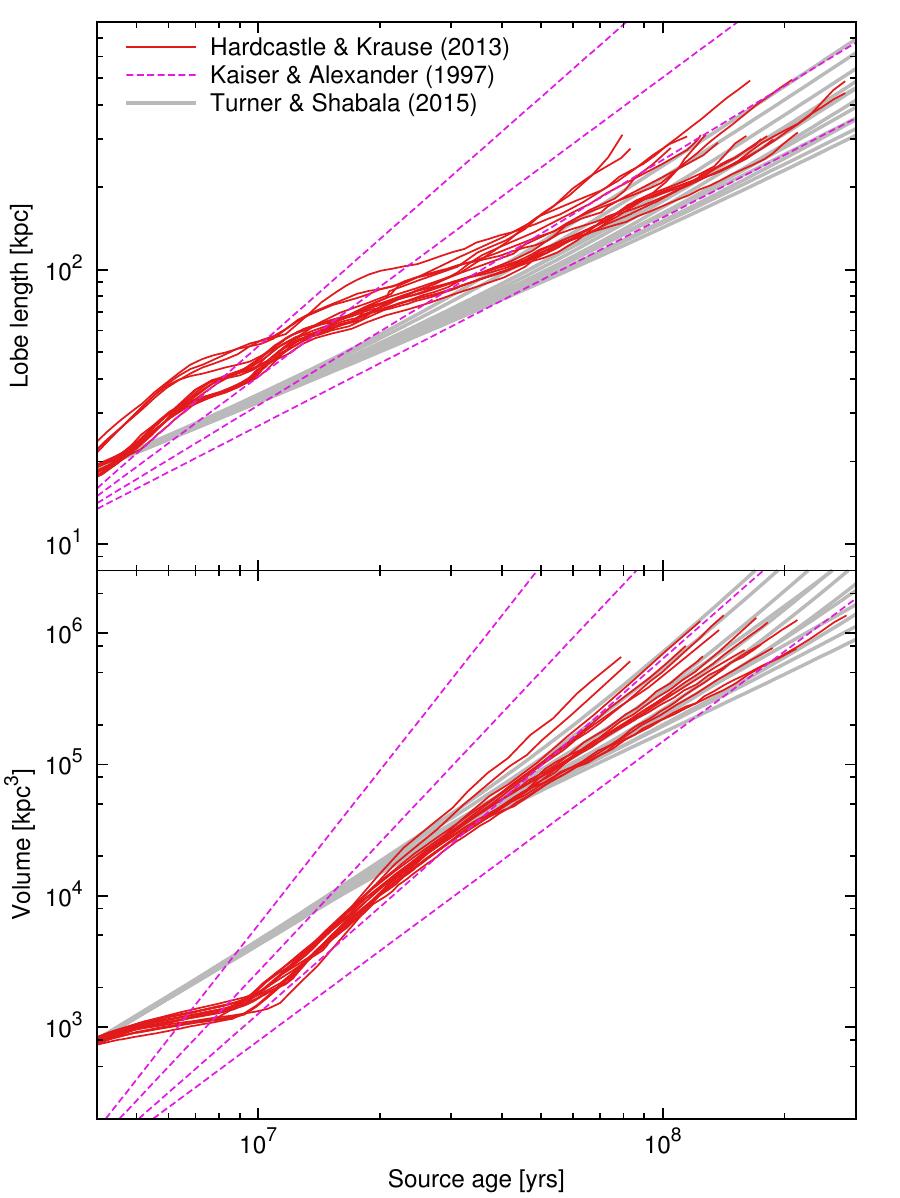} 
\end{center}
\caption{Radio source evolution estimated using our lobed FR-I/II dynamical model compared to that predicted by the hydrodynamical simulations of \citet{HK+2013}. The top panel plots the average length of each lobe as a function of the source age for $10^{38} \rm\, W$ jets expanding into 16 host environments modelled by using different King profiles. The \citet{HK+2013} evolutionary tracks are shown in red with that of the lobed FR-I/II dynamical model plotted in grey for the same parameters. The standard analytic radio source expansion model of \citet{KA+1997} is also shown in pink (dashed) for four comparable environments. The bottom panel plots the lobe volume as a function of source age, noting that the \citet{HK+2013} definition of which particles constitute the lobe may differ from ours.}
\label{fig:dynoplot}
\end{figure}

We follow \citet{HK+2013} in performing hydrodynamical simulations using the PLUTO\footnote{\url{http://plutocode.ph.unito.it/}} code. 
The lobe length and volume evolution as a function of source age is compared in Figure \ref{fig:dynoplot} for radio sources with $10^{38} \rm\, W$ jets expanding into 16 different environments. These host environments are modelled using a King profile \citep[used by][]{HK+2013} with core density $n = 1.5\times10^4 \rm\, m^{-3}$ and temperature $k_{\rm B}T = 1.5\rm\,keV$, but with differing core radii and density fall-off rates. The ages estimated by \citet{HK+2013} are somewhat sensitive to the sound speed of the host environment leading to a horizontal shift in their results in our Figure \ref{fig:dynoplot}.
Their model also assumes a jet half opening angle of $15^\circ$, corresponding to an initial axis ratio of $A = 3$ for relativistic fluid flow \citep{Komissarov+1998}. We find that the lobe length evolution estimated by both models is consistent in all 16 environments, however their lobe volumes are a factor of a few lower at early times (approximately $10 \rm\, Myrs$ for these sources). The spike in the \citet{HK+2013} simulations results from their explicit modelling of the jet before the formation of the radio lobes.

These simulations are further compared to the standard analytic radio source expansion model of \citet{KA+1997}. This model assumes the lobe length expands with a constant power-law dependence as $\mathcal{L} \propto t^{3/(5 - \beta)}$ throughout the radio source evolution, rather than changing with the lobe expansion speed and steepening host environment. Similarly their lobe volume increases with age as $V \propto t^{9/(5 - \beta)}$. The evolutionary tracks predicted using this model therefore diverge considerably from both the lobed FR-I/II dynamical model and the simulations of \citet{HK+2013}. The \citet{Turner+2015} analytic expansion model thus appears to be more representative of lobe dynamics.

\subsubsection{Evolution of lobe axis ratio}
\label{sec:Size evolution of axis ratio}

Most existing dynamical models predict that the axis ratios of powerful FR-II radio sources expand in a self-similar manner, maintaining their initial axis ratio through to the present-time \citep{Leahy+1984, Leahy+1989}.
However, the largest radio sources ($D > 100\rm\,kpc$) are typically more elongated than smaller radio galaxies \citep{Mullin+2008}, suggesting an increase in axis ratio with size. The radio AGNs in this sample are well studied objects with high dynamical range (flux peak\,/\,sensitivity), ensuring the measured axis ratio is not a function of the survey sensitivity limit \citep[c.f.][]{Turner+2016}. The lobe axis ratio is expected to increase in part because the expansion along the longer jet axis works against a lower density environment further from the nucleus. In weaker sources the lobe expansion normal to the jet will slow greatly upon entering the subsonic expansion phase, whilst the supersonic part of the lobe expanding along the jet axis is unaffected \citep[e.g.][]{Alexander+2002, Gaibler+2009, Turner+2015}. The axis ratio of such `transonic' sources can increase rapidly even without the inclusion of additional processes such as Rayleigh-Taylor mixing or Kelvin-Helmholtz instabilities \citep[c.f.][Figure 16]{Gaibler+2009}.

\begin{figure}
\begin{center}
\includegraphics[width=0.48\textwidth]{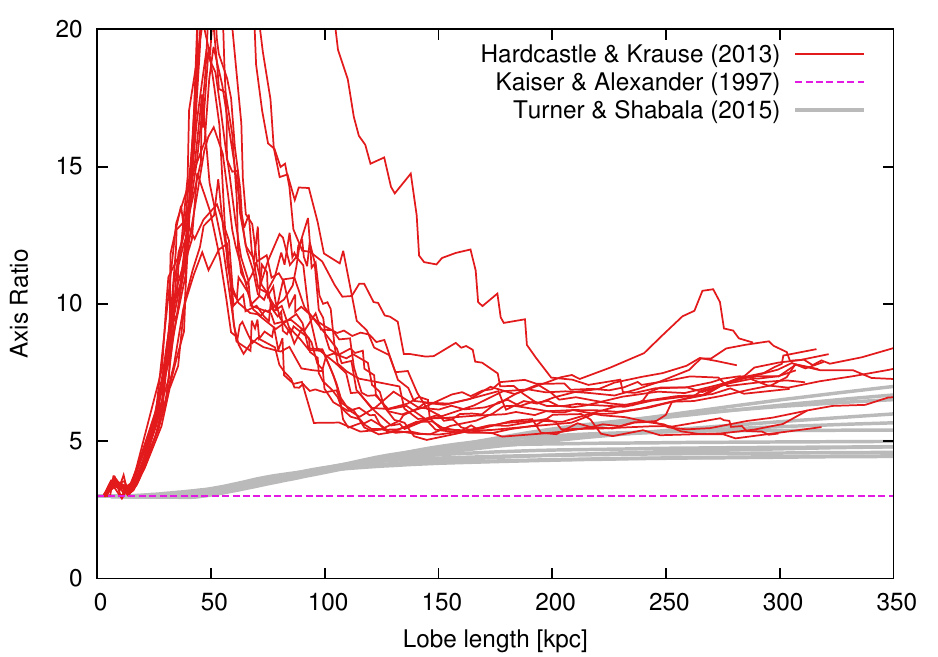} 
\end{center}
\caption{Axis ratio evolution as a function of the lobe length estimated using our dynamical model compared to that predicted by the hydrodynamical simulations of \citet{HK+2013}. All lines are the same as in Figure \ref{fig:dynoplot}. The large spike in axis ratios predicted by \citet{HK+2013} at small sizes is due to the jet punching through the host environment before the lobe has been inflated.}
\label{fig:axisdyno}
\end{figure}

The axis ratio evolution as a function of the lobe length predicted by our dynamical model is compared to that estimated by the hydrodynamical simulations of \citet{HK+2013} in Figure \ref{fig:axisdyno}. The initial axis ratios of our dynamical model and that of \citet{KA+1997} are set to $A = 3$ corresponding to the $15^\circ$ jet opening angle selected by \citet{HK+2013}. Their simulated axis ratios spike at approximately $50\rm\, kpc$ as the jet reaches its maximal length before lobe formation commences, fattening the radio source and decreasing the axis ratio. The lobes form later due to the backflow of plasma from the jet terminal hotspot, reducing the axis ratio back to values similar to those predicted by our lobe dynamical model for larger source sizes. 
{The \citet{Turner+2015} dynamical model and the hydrodynamical simulations thus are in reasonable agreement for larger sources (lobe length $>$$150\rm\, kpc$) and are consistent with an axis ratio that increases with age.}


\begin{figure}
\begin{center}
\includegraphics[width=0.48\textwidth]{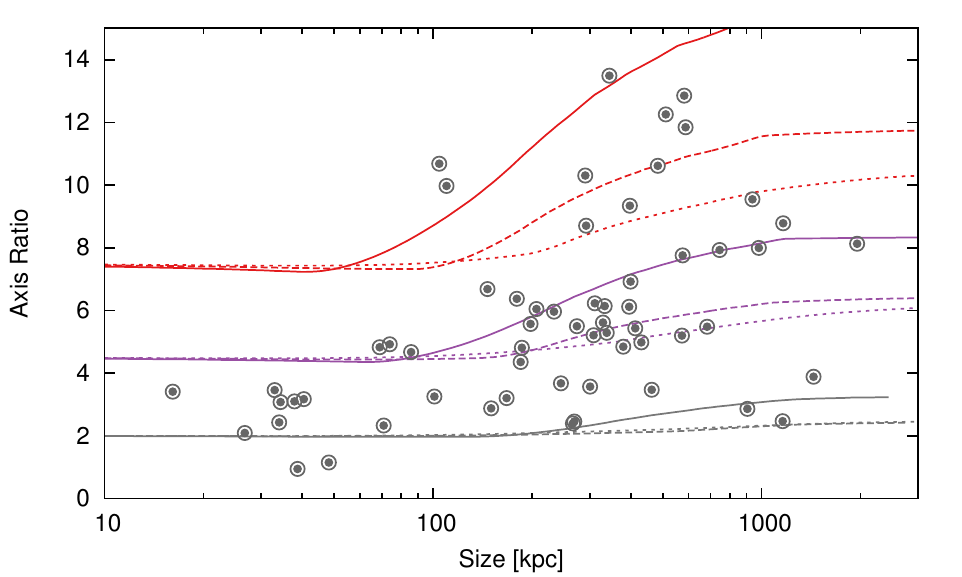} 
\end{center}
\caption{The axis ratio evolution predicted by the lobed FR-I/II dynamical model as a function of source size for a range of jet powers and opening angles (i.e. initial axis ratios). The modelled evolution for the mean initial axis ratio of the 3C sub-sample (purple) and the $\pm1\sigma$ values (grey and red) is plotted assuming typical modelled jet powers. For each initial axis ratio, we plot the axis ratio evolution for the mean host galaxy stellar mass (dashed line), the mean lowered $1\sigma$ (solid line) and raised $1\sigma$ (dotted line). Superimposed on this plot are the observed present-time axis ratios of the sources in our sample. Clearly, using the typical jet powers and environments, the large spread in observed axis ratios can be simulated with a narrow initial range.}
\label{fig:axisevolution}
\end{figure}

We further compare the axis ratio evolution predicted using our lobed FR-I/II dynamical model to that of observed 3C radio sources in Figure \ref{fig:axisevolution}. We simulate this evolution for three host halo environments and initial axis ratios, corresponding to the median and $1\sigma$ uncertainties of our 3C radio source sample. The initial axis ratios are estimated to have a median and $1\sigma$ uncertainty of $A = 4.7_{-2.4}^{+2.9}$. The jet power has a similar but weaker effect on the axis ratio evolution to that of the environment and is omitted from the plot for simplicity. Superimposed on the plot are the observed present-time axis ratios of these sources taken from \citet{Mullin+2008}. The lobed FR-I/II dynamical model reproduces the rapid increase in axis ratio observed for sources larger than $100\rm\,kpc$. However, the \citet{Mullin+2008} observations in isolation cannot distinguish between the possibilities of our hypothesised axis ratio evolution throughout the source lifetime, or some selection effect whereby high initial axis ratio sources tend to be visible at larger sizes (e.g. because they have higher surface brightness).

\subsection{Calibration of integrated luminosity model}

The jet power estimates of \citet{Turner+2015} and similar work \citep[e.g.][]{Shabala+2008} have been made incognisant of the scaling factors in the synchrotron emission model. However, the physics of these objects is not fully understood leading to large uncertainties in the kinetic power-to-luminosity conversion factor. \citet{Willott+1999} described these uncertainties using a single factor $f$ such that $L \propto f^{-3/2} Q^{7/6}$, which is observationally constrained to lie between $\sim 1$ and $20$. There are three main causes of uncertainty in our calculation of the radio luminosity: (1) knowledge of the deviation of the magnetic field strength from equipartition, (2) the energies of the injected synchrotron electrons, and (3) the particle composition of the lobe plasma. The uncertainty in the lobe filling factor considered by \citet{Willott+1999} is directly modelled through the Rayleigh-Taylor mixing of the lobed FR-I/II dynamical model\footnote{{The powerful FR-IIs in the 3C sample are found to have filling factors close to unity; i.e. the transonic expansion of their lobes make them largely impervious to Rayleigh-Taylor mixing.}}. The equipartition factor and lobe composition will be constrained in this section by dynamical model fits to 3C radio sources, whilst constraints on the low-energy cut-off to the electron population will be guided by observations. In this manner, our synchrotron luminosity model can be used to infer the intrinsic properties of AGNs using the luminosity as a constraint.

\subsubsection{Minimum electron Lorentz factor}
\label{sec:Lorentz factor of synchrotron electrons}

The modelled radio source luminosity is moderately sensitive to the low-energy cut-off of the acceleration-time synchrotron electron population. This parameter primarily affects the normalisation of the simulated radio luminosity which we calibrate using observations. It additionally introduces a dependence on the injection spectral index, $\alpha_{\rm inj} = (s - 1)/2$. Specifically, the minimum Lorentz factor of this electron energy distribution is related to the luminosity as $L_{\rm rad} \propto {\gamma_{\rm min}}^{s-2}$ \citep[Equation 8 of ][]{KDA+1997}. The minimum Lorentz factor in the hotspot has been estimated by other authors for selected radio sources including: Cygnus A \citep{McKean+2016}, PKS\,1421-490 \citep{Godfrey+2009}, 3C123, 3C196 and 3C295 \citep{Hardcastle+2001}. These hotspot Lorentz factor estimates are mostly in the order of a few hundred {although these estimates are quite uncertain due to the confounding effects of free-free and synchrotron self-absorption}. The Lorentz factor of the electron population in the lobe is expected be approximately an order of magnitude lower due to the adiabatic expansion of the plasma. The \citet{Turner+2016} lossy luminosity model used in this paper, based on the work of \citet{KDA+1997}, simulates the adiabatic expansion of the packets of synchrotron electrons from the higher pressure hotspot to the lobe at the time of emission. The minimum Lorentz factor of the electron energy distribution at the hotspot reduces accordingly as the electrons propagate throughout the lobe; the Lorentz factor in the lobe does not need to be explicitly measured. 
Following \citet{Godfrey+2009} we assume a typical hotspot minimum Lorentz factor of $\gamma_{\rm min} = 500$ for the sources in our sample. Systematic errors in AGN intrinsic parameter estimates are derived for realistic Lorentz factors in the range $\gamma_{\rm min} = 100$-$1000$, yielding a $0.19\rm\,dex$ uncertainty in jet power estimates, $0.06\rm\,dex$ in source ages, and $0.01\rm\,dex$ in equipartition factors.
This observationally informed choice should ensure a realistic luminosity dependence on the injection index, whilst the normalisation remains to be fully calibrated.

\subsubsection{Lobe magnetic field strength}
\label{sec:Lobe equipartition factor}

The scaling and shape of the radio spectrum depend on the magnetic field strength; the fraction of energy stored in the magnetic field must therefore be quantified in order to constrain the AGN kinematics. This energy fraction is considered here as the deviation of the field strength $B$ from the equipartition value $B_{\rm eq}$, in which equal energy is assumed to be in both the field and the synchrotron-emitting electrons. This equipartition factor is related to the {ratio of energy density in the magnetic field to that in the synchrotron-emitting particles}, $q$, through $B/B_{\rm eq} = q^{2/(s + 5)}$ \citep[e.g.][]{Croston+2005}. Below, we use the radio lobe dynamics alone to show that the 3C radio sources cannot be in equipartition, {then use further observations to precisely constrain the magnetic field strength}.

\begin{figure}
\begin{center}
\includegraphics[width=0.48\textwidth]{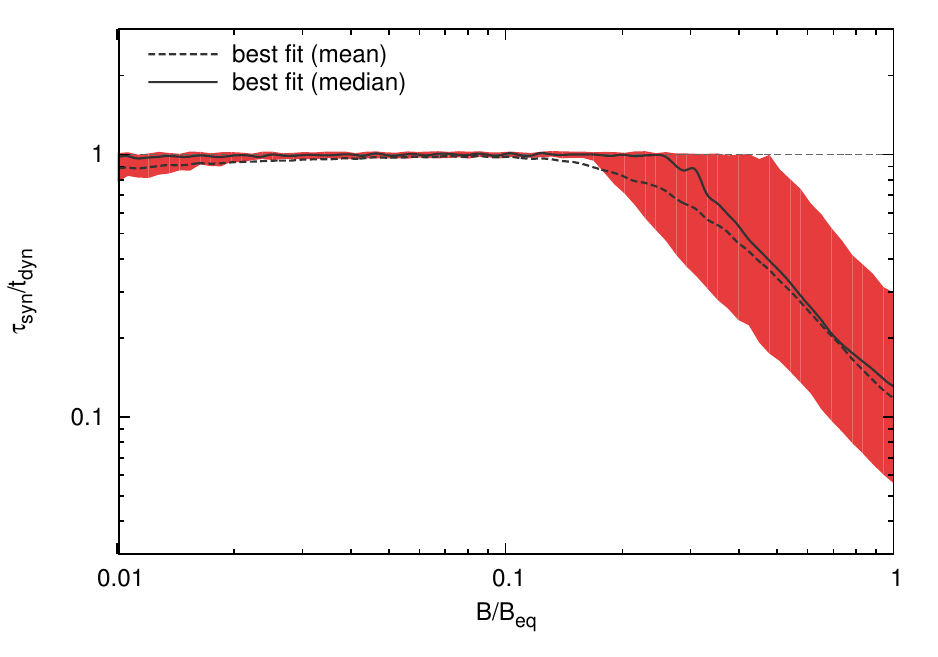} 
\end{center}
\caption{Ratio of fitted synchrotron and dynamical ages as a function of the deviation from the equipartition magnetic field strength. The black line plots the mean ratio for a reduced sample of 37 3C sources with sufficient data. The median ratio is plotted in red with the $1\sigma$ uncertainties is shown by the shading. Radio lobes close to equipartition have unrealistically young spectral age estimates which can not be reproduced by our dynamical model for these sources.}
\label{fig:equifitplot}
\end{figure}

{A Bayesian fitting algorithm (see Section \ref{sec:Bayesian parameter estimation}) is used to determine the most likely dynamical and spectral age for 37 3C radio sources based on their observed size and spectral break frequency. The ages are fitted assuming a range of magnetic field strengths (or equivalently equipartition factors $B/B_{\rm eq}$) and halo environments taken from the semi-analytic SAGE model for their observed stellar mass.} For each equipartition factor we determine the difference between the best fit spectral and dynamical age for each source, with the median and $1\sigma$ of the distribution shown in Figure \ref{fig:equifitplot}. The dynamical and spectral ages are inconsistent for all of our sources if we assume the lobes are in equipartition. By contrast, the two age estimates agree if the 3C sample has a median equipartition factor of $B/B_{\rm eq} \lesssim 0.3$, corresponding to an energy density ratio of $q \lesssim 0.01$. These upper bounds are derived entirely using the radio source dynamics and fits to the spectral curvature, and thus are not affected by any uncertainties associated with the power-to-luminosity conversion factor.

\begin{figure}
\begin{center}
\includegraphics[width=0.48\textwidth]{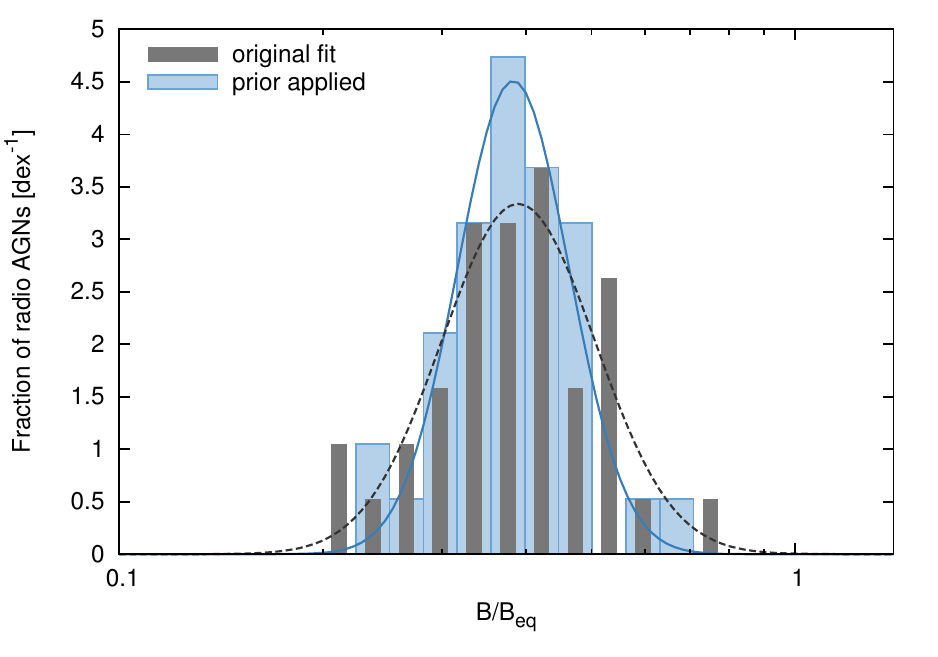}
\end{center}
\caption{Fitted distribution for the equipartition factor using the {\sc Full\,Fit} algorithm (see Section \ref{sec:PARAMETER ESTIMATION}), both with (blue) and without (grey) the use of the prior probability distributions. The solid and dashed lines are Gaussian fits to these probability distributions respectively.}
\label{fig:equidist_plot}
\end{figure}

The distribution of equipartition factors for these 37 3C sources can be quantified using the calibrated monochromatic luminosity as an additional constraint. The resulting distribution shown in Figure \ref{fig:equidist_plot} is approximately normal having a median of $B/B_{\rm eq} = 0.38$, $1\sigma$ point uncertainty of $0.17 \rm\, dex$, and standard error of $0.10 \rm\, dex$ (including systematic uncertainties from the environment and electron population). The lobe magnetic field strengths of the sources in the reduced 3C sample are thus lower than the equipartition value by a factor of three (significant at the $5\sigma$ level). 
The stability of the parameter fitting algorithm is tested by using the original distribution as a prior probability distribution instead of a flat prior. This choice of prior does not affect our conclusions, as shown in Figure \ref{fig:equidist_plot}.

\begin{figure}
\begin{center}
\includegraphics[width=0.42\textwidth]{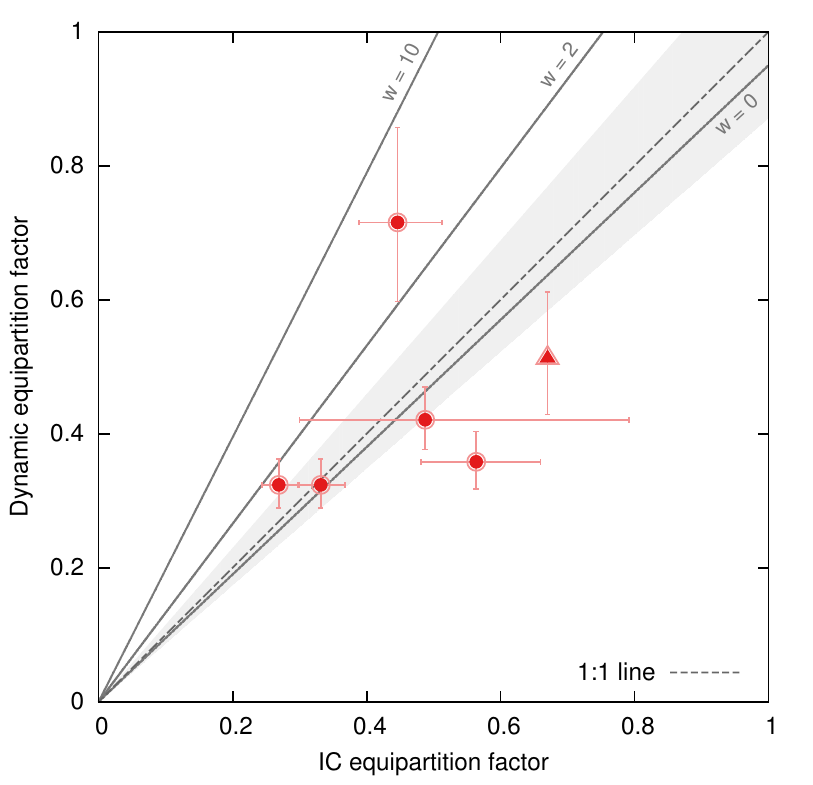}
\end{center}
\caption{Estimated equipartition factor $B/B_{\rm eq}$ for 3C219N, 3C219S, 3C244.1, 3C330, 3C427.1 and Pictor A as a function of the X-ray inverse-Compton (IC) measurements of \citet[][triangle]{Hardcastle+2016} and \citet[][circles]{Ineson+2017}. The six unique measurements of the field strength for the 3C sources are plotted assuming $w = 0$ with $1\sigma$ errorbars included for each data point except Pictor A (triangle at top). {The best linear fits to this and equivalent distributions assuming alternative proton-to-lepton energy density ratios (of $w = 2$ and 10) are shown as solid lines}, with the $1\sigma$ standard errors in the fits shaded and centred on the one-to-one line.}
\label{fig:eqeq_plot}
\end{figure}

These equipartition factors are consistent with the observations of other authors. \citet{Croston+2005} and \citet{Ineson+2017} found the energy density in powerful FR-IIs peaks at $B/B_{\rm eq} \sim 0.5$ (corrected to observed spectral index $s = 2\alpha_{\rm inj} + 1$) and 0.4 respectively based on X-ray inverse-Compton emission from the synchrotron emitting electron population. {The dynamical model based equipartition factors are also in good agreement with the 3C observations of \citet{Ineson+2017} on a source-by-source basis. These measurements are all consistent to within the $2\sigma$ uncertainties, as shown in Figure \ref{fig:eqeq_plot}. We note that the \citet{Ineson+2017} equipartition factors are calculated assuming an injection index of $s = 2.4$, in contrast to the $s=2.05$-$2.6$ range fitted to the 3C spectra; we have therefore adjusted their measurements for this variation in injection index.} \citet{Hardcastle+2014} similarly find that radio synchrotron and inverse Compton emission properties derived from substantially sub-equipartition magnetohydrodynamic (MHD) simulations are broadly in good agreement with observations. 
The general consensus is thus of FR-II lobes with field strengths approximately a factor two to three less than the equipartition value, or equivalently magnetic field energy densities approximately a factor of {40} less than that in the electrons.

The similarity between our dynamical model equipartition factor estimates and those of \citet{Ineson+2017} is encouraging, {however all but a single source (3C427.1) have dynamical equipartition factors fitted in the narrow range $0.32 < B/B_{\rm eq} < 0.42$}. The veracity of our fitting algorithm is further tested against observations of Pictor A, which has a measured equipartition factor of $\sim$$0.7$ \citep{Hardcastle+2016}. The spectrum of Pictor A is fitted using our continuous injection model (see Section \ref{sec:SPECTRAL EVOLUTION MODELS}) based on data taken from \citet{Kuehr+1981}, whilst other properties are taken from \citet{Hardcastle+2016}. The equipartition factor thus derived for Pictor A is $B/B_{\rm eq} = 0.51\pm0.07$. The distribution of equipartition factors estimated for the 3C sources is therefore not an artefact of the fitting algorithm; our method appears capable of accurately determining the magnetic field strength in close-to-equipartition sources.


\subsubsection{Radio lobe plasma composition}
\label{sec:Radio lobe plasma composition}

{The particle composition of the lobes of radio AGN is arguably most directly probed through X-ray observations of the inverse-Compton emission; although incapable of directly detecting any non-radiating particles, observations of close-to-equipartition lobes are used to suggest at most small fraction of the energy density is present in relativistic protons or a thermal plasma \citep[e.g.][]{Hardcastle+2002, Croston+2004, Croston+2005, Ineson+2017}}. 
In this technique the X-ray emission upscattered from the synchrotron-emitting electrons is compared to that expected if their energy densities are in equipartition with the magnetic field. 
This direct measurement of the equipartition factor is therefore unaffected by the inclusion of {non-radiating} particles into the lobe plasma. {By contrast, the dynamical model measurements are dependent upon the lobe plasma composition;} the equipartition factors derived using our model for 3C219, 3C244.1, 3C330, 3C427.1 and Pictor A are thus recalculated for a range of proton-to-lepton {(or non-radiating to radiating particle)} energy density ratios, as shown {by the fitted lines} in Figure \ref{fig:eqeq_plot}. The inclusion of {non-radiating} particles does permit the possibility of radio lobes in equipartition, although the energy density in thermal particles is required to be up to 35 times that of the leptons. By contrast, proton-to-lepton energy density ratios between $w = 0$ and $2$ are consistent with the equipartition factors measured by \citet{Hardcastle+2016} and \citet{Ineson+2017} to within the $2\sigma$ uncertainties.

\begin{figure}
\begin{center}
\includegraphics[width=0.42\textwidth]{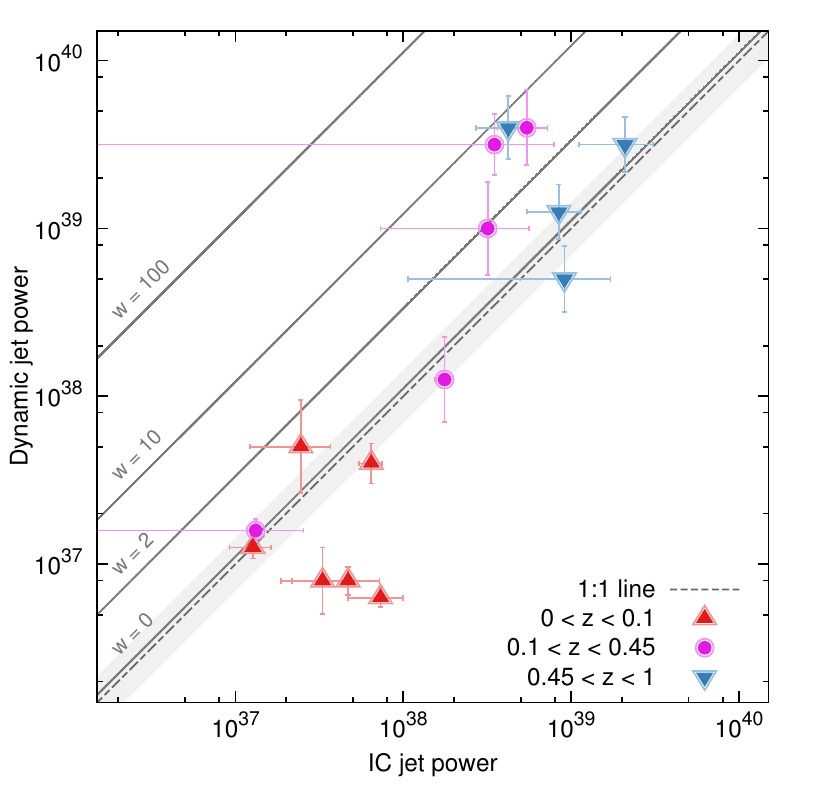}
\end{center}
\caption{Correlation between our dynamical model based jet power estimates (either {\sc Full\,Fit} or {\sc Source\,Size} fitting algorithm, see Section \ref{sec:PARAMETER ESTIMATION}) and the \citet{Ineson+2017} powers calculated from X-ray inverse-Compton measurements. The apparent strength of the correlation results largely from the redshift-dependent flux-detection limits. Radio sources at $z < 0.1$ are shown in red (triangles), those at $z > 0.45$ in blue (inverted triangles), with the remainder plotted in pink (circles). The jet powers for these plotted points are calculated assuming $w = 0$. 
{The best linear fits to this and equivalent distributions assuming alternative proton-to-lepton energy density ratios (of $w = 2$, 10 and 100) are shown as solid lines}, with the $1\sigma$ standard errors in the fits shaded and centred on the one-to-one line.}
\label{fig:QQplot2}
\end{figure}

The energy density in relativistic protons or a thermal plasma is tested further by comparing the jet powers estimated for our 3C sample (see Section \ref{sec:PARAMETER ESTIMATION}, and {Tables \ref{tab:fittedvalues} and \ref{tab:fittedvalues2}}) to independent measurements. {\citet{Ineson+2017} use their inverse-Compton modelling to derive jet power estimates from their measurements of the lobe internal pressure and its expansion speed (which provides an age estimate).} The correlation between these jet power measurements and our dynamical model based estimates is shown in Figure \ref{fig:QQplot2}. However, this strong relationship results primarily from the redshift dependent nature of the survey detection limits. Examining only radio sources from a small redshift range may therefore yield largely scatter (Figure \ref{fig:QQplot2}), as shown conclusively by \citet{Godfrey+2016} for a range of jet power measurement techniques.
Regardless, the results in Figure \ref{fig:QQplot2} provide some confidence in the jet power-to-luminosity conversion factor of our model and support the previous conclusions about the composition of the lobe plasma.
Radio sources with lobe plasma {exclusively comprising radiating particles} have jet powers consistent with the X-ray inverse-Compton measurements. By contrast, a lobe comprising an energetically dominant population of relativistic protons or thermal plasma (i.e. $w > 2$) is inconsistent with our estimates at the $2\sigma$ level.

The comparison of our dynamical model based jet power and equipartition factor estimates to independent measurements from \citet{Hardcastle+2016} and \citet{Ineson+2017} have yielded the same outcome: a proton-to-lepton ratio of $w < 2$ at the $2\sigma$ level. There is therefore strong evidence against an energetically dominant population of relativistic protons or thermal plasma (i.e. $w > 2$). These results are entirely consistent with a lobe excluding any significant energy in a {population of non-radiating particles}, although based on these findings $w \lesssim 2$ is quite plausible.  
In this work, we will therefore adopt a pair-plasma with no energetically-dominant proton population (i.e. $w = 0$). The possibility of a population of {non-radiating particles} with up to double the energy of that in the leptons cannot be excluded, however our results are not greatly affected for $w \lesssim 2$.

\section{PARAMETER ESTIMATION}
\label{sec:PARAMETER ESTIMATION}

{The intrinsic AGN parameters, including the kinetic jet power, source age and magnetic field strength, are derived from the CI model fits to the radio spectra and other observables using a Bayesian approach}. 
In this section, we consider various fitting algorithms utilising either all the observed parameters or only a limited set that might be expected to be available in large-scale surveys. 
The versatility of the AGN parameter fitting {technique} potentially enables intrinsic properties {to be estimated for} high-redshift, unresolved sources {for which no size constraint can be determined}.

\subsection{Bayesian parameter estimation}
\label{sec:Bayesian parameter estimation}

The most likely jet powers $Q$, source ages $t$, and equipartition factors $B/B_{\rm eq}$ are estimated for our sample of 3C radio sources using a Bayesian approach. {These intrinsic parameters are constrained based on the observed luminosity, spectral break frequency and source size, and the host galaxy environment (see Section \ref{sec:SIMULATED AGN HOST ENVIRONMENTS})}. 
The {fitting} algorithms tested here include: utilising all the observables ({\sc Full\,Fit}); size and calibrated monochromatic luminosity ({\sc Source\,Size}); spectral break frequency and luminosity ({\sc Spectral\,Break}); and using only the monochromatic luminosity at $178\rm\,MHz$ ({\sc Luminosity\,Only}). 
We simulate the radio source evolution for a range of jet powers, ages and equipartition factors assuming typical environments for each source's stellar mass and redshift. The goodness of fit of these evolutionary models for each age/jet power/equipartition factor triplet to some observable $x$ is assessed using the chi-squared probability 

\begin{equation}
P(x) = \exp \left(- \left[\frac{\log x_{\rm mod} - \log x_{\rm obs}}{\sigma_{\log x}} \right]^2 \middle/ 2 \right) ,
\end{equation}

where $\log x_{\rm mod} - \log x_{\rm obs}$ is the log-space difference between the observed and simulated values of parameter $x$. 
The uncertainty due to the model dominates and is taken as $\sigma_{\log x} = 0.3 \rm\, dex$, noting {that any values close to this} yield stability in the solution space. The parameters used for each fitting algorithm are listed in Table \ref{tab:fitting}. 
The likelihood of each simulation matching the observations is calculated as the product of the probabilities for each parameter, with the age/jet power/equipartition factor triplet associated with the maximum likelihood taken as our best estimates.

\begin{table}
\begin{center}
\caption[]{Bayesian fitting algorithms used to derive the jet powers, source ages, sizes and equipartition factors for the 3C radio source sub-sample. These algorithms use the probability of the evolutionary track matching the observed size $P_{\rm\:\! D}$, spectral break frequency $P_{\rm \nu_{\rm b}}$, and luminosity $P_{\rm\:\! L}$, in various combinations. In the final column, the number of 3C objects with the requisite data for each algorithm is shown.}
\label{tab:fitting}
\setlength{\tabcolsep}{10pt}
\begin{tabular}{ccccc}
\hline\hline \vspace{-0.00cm}
Algorithm Name&\!$P_{\rm\:\! D}$\!&\!$P_{\rm \nu_{\rm b}}$\!&\!$P_{\rm\:\! L}$\!&\!$n_{\rm obs}$\! \vspace{0.000cm}\\
\hline \vspace{-0.00cm}
\textsc{Full Fit}&yes&yes&yes&37\\
\textsc{Source Size}&yes&-&yes&71\\
\textsc{Spectral Break}&-&yes&yes&37\\
\textsc{Luminosity Only}&-&-&yes&71\vspace{0.000cm}\\
\hline
\end{tabular}
\end{center}
\end{table}

The radio source evolution is additionally affected by the jet opening angle which in turn affects the axis ratio of the young lobe \citep{Komissarov+1998}. We model the evolution of our sources for a range of initial axis ratios and estimate the most likely ratio at the time of observation based on our source age, jet power and equipartition factor estimates. The initial axis ratio yielding a present-time ratio closest to the observed value from \citet{Mullin+2008} is taken as our best estimate.

\subsection{Ill-conditioned fitting algorithms}
\label{sec:Ill-conditioned fitting algorithms}

The {\sc Source\,Size}, {\sc Spectral\,Break} and {\sc Luminosity\,Only} fitting algorithms are underdetermined, potentially leading to unstable solutions. However, rather than yielding no solution, these algorithms derive some parameters robustly whilst others are poorly constrained. As we explain in Appendix \ref{sec:Ill-conditioned appendix}, the source age and lobe pressure are degenerate variables, enabling the omission of the parallel size constraint in the {\sc Spectral\,Break} algorithm. The jet power and equipartition factor are also somewhat degenerate yielding reliable source age estimates using the {\sc Source\,Size} algorithm. The intrinsic parameters for the 37 radio sources with a complete set of observational constraints are included in Table \ref{tab:fittedvalues}. In this section, the stability of the underdetermined algorithms is directly compared to the results of the full model. Prior probability distributions are also applied during the fitting process to yield more stable results for some algorithms.

\subsubsection{Size--luminosity fitting algorithm}
\label{sec:Size--luminosity fitting algorithm}

{The \citet{Mullin+2008} 3C sample could only have spectral break frequencies confidently fitted to $52\%$ of radio sources, after the removal of AGNs with bright cores, peculiar spectra or no observations of their spectra}. Further, in weaker sources than observed here, the low signal-to-ratio noise may preclude an accurate fit of the break frequency. There may therefore be numerous instances where using the spectral break frequency in the fitting algorithm is not viable, but the source size is known. However, by excluding the break frequency from the fit we are no longer capable of constraining all free parameters, and a typical value must be assumed for the lobe equipartition factor, e.g. $B/B_{\rm eq} = 0.38$ (Section \ref{sec:Source age--lobe pressure degeneracy}). This fitting method was used in \citet{Turner+2015} and by others including \citet{Shabala+2008}.

\begin{figure}
\begin{center}
\includegraphics[width=0.48\textwidth]{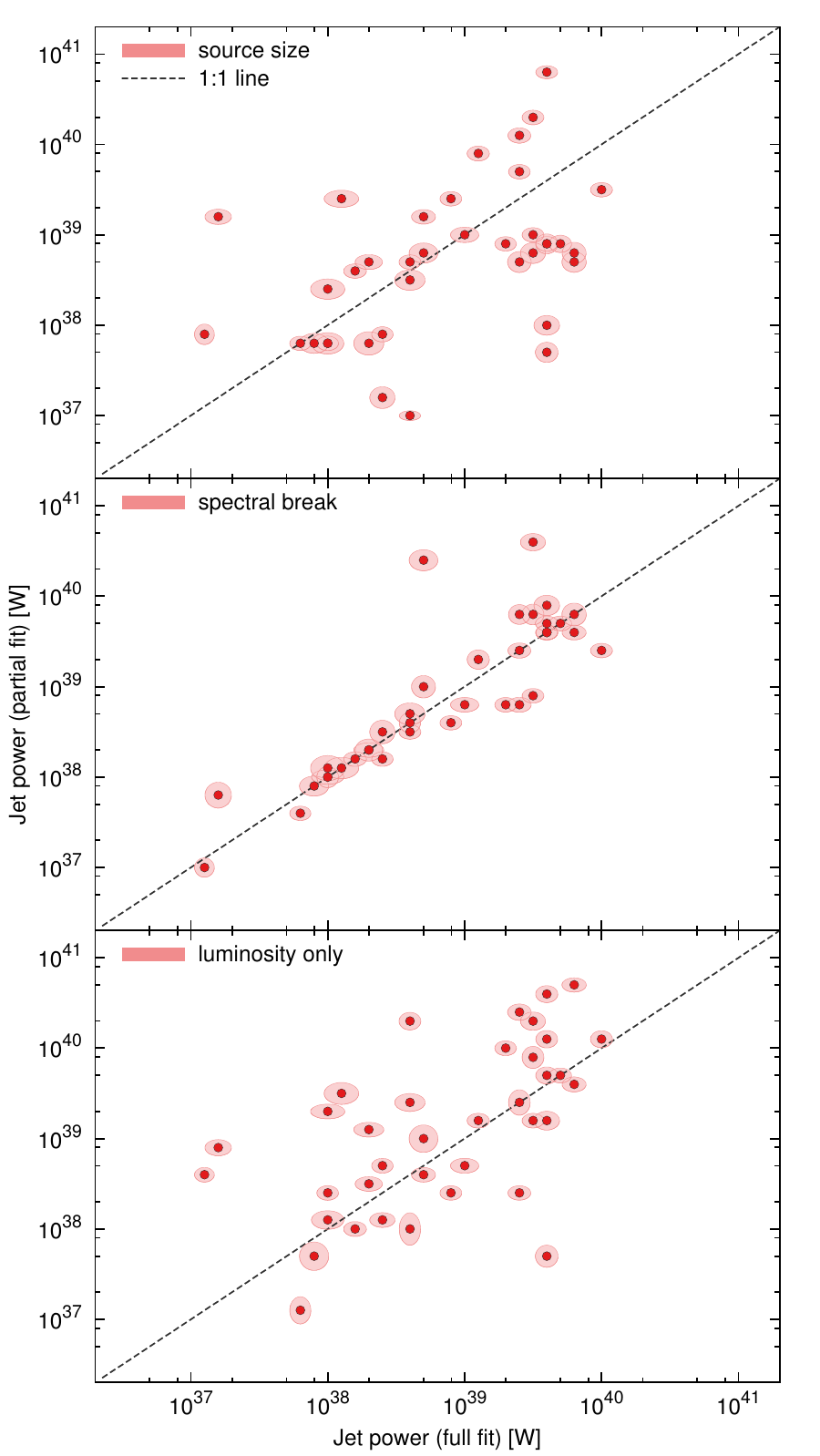}
\end{center}
\caption{The consistency of jet power estimates is compared to the best estimates for fitting algorithms using the two observables size and luminosity (top panel), break frequency and luminosity (middle panel), and a single variable fit using only luminosity (bottom panel). The jet powers estimated using the three variable full fit are plotted on the horizontal axis with the jet powers of the simpler models on the vertical axis. The red circles and associated ellipses show the mark the $1\sigma$ model uncertainties for each radio source with the dashed line marking a one-to-one relationship.}
\label{fig:QQplot}
\end{figure}

\begin{figure}
\begin{center}
\includegraphics[width=0.48\textwidth]{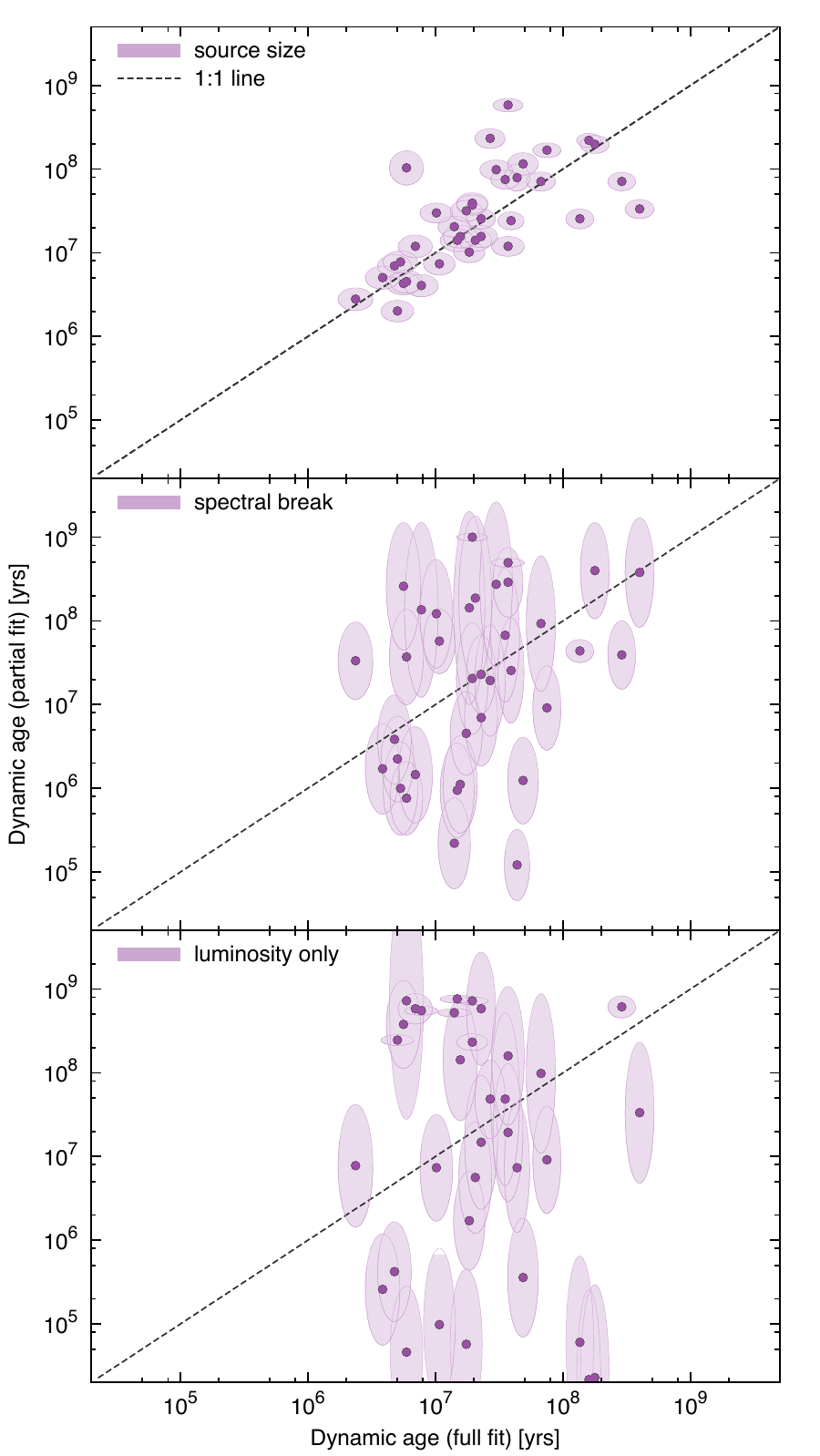}
\end{center}
\caption{Same as Figure \ref{fig:QQplot} but comparing the consistency of the source age estimates to the full fitting algorithm for each model simplification.}
\label{fig:ttplot}
\end{figure}

\begin{figure}
\begin{center}
\includegraphics[width=0.48\textwidth]{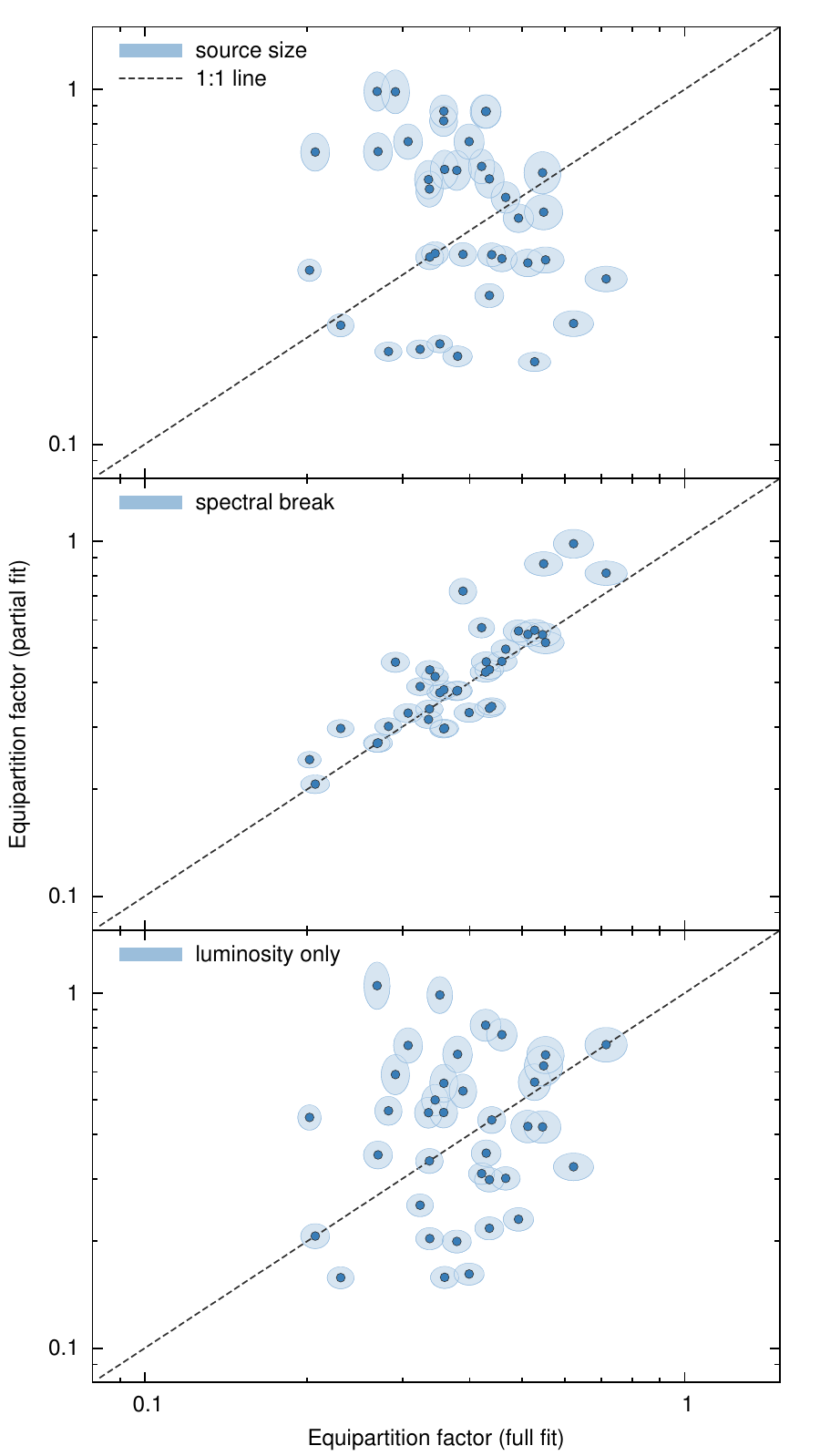}
\end{center}
\caption{Same as Figure \ref{fig:QQplot} but comparing the consistency of the equipartition factor estimates to the full fitting algorithm for each model simplification.}
\label{fig:eqeqplot}
\end{figure}

The stability of the {\sc Full\,Fit} algorithm to the removal of the spectral break frequency parameter is examined in the top panel of Figures \ref{fig:QQplot}, \ref{fig:ttplot} and \ref{fig:eqeqplot}. Without the use of a prior probability distribution for the equipartition factor the jet powers estimated using the {\sc Source\,Size} algorithm differ considerably from the full model ($r^2 = 0.20$). The inclusion of the equipartition factor prior (not shown in figure) enables the {\sc Source\,Size} fitting algorithm to explain $75\%$ of the variation in the jet powers estimated by the full model. 
The source ages are only marginally better recovered than the jet powers without the use of an equipartition prior, with only $46\%$ of the variation in the full fit explained. This is increased to $87\%$ when the equipartition factor prior is assumed. However, the {\sc Source\,Size} algorithm is incapable of estimating the equipartition factor ($r^2 < 0.05$), with the use of priors simply yielding the peak value of the equipartition factor prior probability distribution. This is consistent with the analytical expectations discussed in Section \ref{sec:Source age--lobe pressure degeneracy}.

The jet power and age estimates from the simpler size--luminosity method are very consistent with those of the full fitting algorithm with the use of a prior for the equipartition factor. This simplification is likely quite reasonable in any case where the break frequency cannot be measured. The jet powers of the 34 3C radio sources without break frequency measurements are therefore calculated using this method and included in Table \ref{tab:fittedvalues2}.

\subsubsection{Break frequency--luminosity fitting algorithm}
\label{sec:Break frequency--luminosity fitting algorithm}

The resolution of the radio lobes for all but the largest sources may not be possible at high-redshifts even with future all-sky surveys; for example, a 10\,arcsec beam FWHM corresponds to $80\rm\,kpc$ limit at $z\sim1$ for ASKAP EMU \citep{Norris+2011}. The ability to understand the effect AGNs have on galaxy formation and evolution in the early universe is thus hindered severely if size is required to constrain their intrinsic properties. Developing a stable fitting algorithm excluding the source size would aid greatly further understanding AGN feedback at this epoch. 

The stability of the {\sc Full\,Fit} algorithm to the removal of the size parameter can now be examined (centre panel of Figures \ref{fig:QQplot}, \ref{fig:ttplot} and \ref{fig:eqeqplot}). The jet power estimates are quite consistent with those estimated using the full model for all but two sources (3C244.1 and 3C247), with $88\%$ of the variation explained by the simpler model (excluding size) if these two poorly fitted sources are excluded from the statistic. The equipartition factors are similarly well reproduced by the simpler fitting algorithm ($r^2 = 69\%$). The adoption of an equipartition factor prior slightly reduces the strength of these correlations. By contrast, the {\sc Spectral\,Break} fitting algorithm is completely incapable of estimating the source ages ($r^2 < 0.05$) due to the degeneracy between the age and lobe pressure (see Section \ref{sec:Source age--lobe pressure degeneracy}). 
The source age estimates are improved for most sources by assuming a prior probability distribution for the lobe pressure based on empirical relationships between observed and intrinsic parameters; e.g. lobe pressure--luminosity. However, this approach is not physically based and can worsen the estimates for some less typical sources by several orders of magnitude; such prior probability distributions are not examined in this work.  

The {\sc Spectral\,Break} fitting algorithm is therefore capable of reliably estimating the jet powers of unresolved radio sources, although it requires size information or at least an upper bound to determine their ages. New radio surveys may not observe at more than two frequencies, and hence it may not be possible to determine the break frequency for all radio sources. This problem should be resolved by combining observations from multiple surveys, with coverage at $1.4\rm\,GHz$ (e.g. ASKAP EMU) or lower frequencies (e.g. MWA GLEAM/SKA-low, LOFAR, VLSS). Break frequencies above $1.4\rm\, GHz$ may require follow-up observations to these surveys at higher frequencies.
Further, the break frequency may not be able to be observed at all for radio sources younger than approximately $1\rm\, Myr$ where the spectral break will be at very high frequencies (e.g. $>$$20\rm\, GHz$). Very young sources will additionally exhibit curvature at lower frequencies due to free-free and synchrotron or self-absorption.

\subsubsection{Luminosity-only fitting algorithm}
\label{sec:Luminosity only fitting algorithm}

Numerous authors have attempted to predict the jet powers of radio sources based solely on their luminosities \citep[jet power--luminosity relationship, e.g.][]{Willott+1999}. However, these empirical relationships are significantly weakened when corrected for the confounding distance variable \citep{Godfrey+2016}. The validity of such methods are further examined here by fitting the 3C sources which have a complete set of constraints using solely the monochromatic luminosity. The stability of the method is shown in the bottom panel of Figures \ref{fig:QQplot}, \ref{fig:ttplot} and \ref{fig:eqeqplot}. 
The jet powers shown are weakly related to those estimated using the full algorithm ($r^2 = 0.34$), though with the jet powers of many sources differing by over an order of magnitude. This is because radio sources undergo significant luminosity evolution \citep{Shabala+2013} whilst the environment also plays an important role \citep{KDA+1997, HK+2013, Turner+2015}. The adoption of an equipartition factor prior probability distribution improves the relationship ($r^2 = 0.60$), though this algorithm is still far worse at estimating jet powers than with the inclusion of a size or spectral break measurements. By contrast, both the age and size estimates show very little resemblance to those calculated using the full algorithm ($r^2 < 0.05$). These jet power--luminosity relationships are therefore acceptable only for an order-of-magnitude estimate of the jet power even if priors can be assumed, whilst using the luminosity on its own as a fit constraint for the age and size is unviable. This is not surprising given the large luminosity evolution experienced by sources over their lifetimes \citep[e.g.][]{Turner+2015}.

\section{CONCLUSIONS}
\label{sec:CONCLUSIONS}

We have presented a new method to measure the AGN jet kinetic power, radio source age and lobe magnetic field strength based on the observed radio lobe luminosity and length, the shape of the AGN synchrotron emission spectrum, and the properties of their host halo environment. Our approach differs in two important respects from previous jet power estimation methods: (1) all parameters in the synchrotron emission model are either fitted or constrained using observations; and (2) confounding variables such as the lobe length and host halo environment are considered. Specifically, the low-frequency cut-off to the synchrotron electron energy distribution is observationally constrained, whilst the filling factor (model dependent) and any departure from the equipartition magnetic field strength are fitted using a Bayesian approach. \citet{Turner+2015} and \citet{Shabala+2008} also consider this size and environment dependence, but do not constrain the uncertainties in the luminosity model or consider the full radio spectra.

The dynamics predicted by our lobed FR-I/II evolution model is found to closely match hydrodynamical simulation results \citep{HK+2013, Hardcastle+2014}. By contrast, the existing self-similar analytic model of \citet{KA+1997} yields results which are largely inconsistent with both our dynamical model and the simulations, mainly because of the the simplified modelling of the host environments. Further, the axis ratio evolution predicted by our dynamical model for the 3C sub-sample is capable of reproducing the present-time axis ratios from a narrow range of initial axis ratios and host halo environments. 

We applied the combined dynamical and synchrotron luminosity model to a sample of 71 3C \citeauthor{FR+1974} type II radio galaxies to infer the jet powers, ages and equipartition factors based on combinations of their observed sizes, luminosities and spectral shapes (where available). The synchrotron spectrum of a subset comprising 37 of these objects was fitted using a continuous injection model, and curvature was parametrised with the break frequency parameter. We tested the sensitivity of estimated AGN physical parameters to availability of data and found:

\begin{enumerate}
  \setlength\itemsep{0.0em}
  \item The full set of observables (break frequency, lobe length and luminosity) allow accurate inference of the jet power, source age and lobe equipartition factor.
  \item Excluding the break frequency (e.g. observing only a low-frequency monochromatic luminosity) allows the source age to be constrained, but {not} the jet power or the equipartition factor.
  \item The spectral break frequency and lobe luminosity accurately infer the jet power and lobe equipartition factor, but not the source age.
  \item Luminosity alone poorly estimates the jet power, and is incapable of inferring the source age and equipartition factor.
\end{enumerate}

We find that the average energy in the lobe magnetic field is approximately a factor of {40} lower than the energy in the particles, or equivalently the lobe field strength is $B/B_{\rm eq} = 0.38$ lower than the equipartition value (significant at the $5\sigma$ level). These findings are consistent with the X-ray inverse-Compton observations of \citet{Ineson+2017} with higher equipartition factors of $B/B_{\rm eq} \sim 0.4$ for a similar sample of powerful FR-II radio sources. 
Radio lobes comprising a pair-plasma with {at most a small pressure contribution from a population of relativistic protons or heated thermal plasma} (i.e. {a ratio of the energy density in the non-radiating particles to that in the leptons of} $w = u_{\rm t}/u_{\rm e} \lesssim 2$) are found to yield results consistent with both the X-ray inverse-Compton measurements of the equipartition factor and jet powers measurements of \citet{Ineson+2017}.

The parameters used in the Bayesian fitting algorithm may be further varied or optimised to explore currently poorly understood objects. The development of versatile tools for estimating source ages and jet powers is especially pertinent with the large volumes of data expected from the \emph{Square Kilometre Array} pathfinder surveys \citep[e.g. ASKAP EMU and MeerKAT MIGHTEE;][]{Norris+2011, Jarvis+2012}. In particular, we have shown that the spectral break frequency and lobe luminosity alone can estimate jet power without the size constraint, enabling both compact and high-redshift sources to be probed.

\subparagraph{}
R.T. thanks the University of Tasmania for an Elite Research Scholarship. S.S. and M.K. thank the Australian Research Council for an Early Career Fellowship, DE130101399. M.K. further thanks the University of Tasmania for a Visiting Fellowship. We thank Prof. Martin Hardcastle and the referee for their helpful and insightful comments which have greatly improved our manuscript.

\begin{appendix}
\section{Parameter degeneracies}
\label{sec:Ill-conditioned appendix}

\subsection{Source age--lobe pressure degeneracy}
\label{sec:Source age--lobe pressure degeneracy}

The spectral break frequency constraint of Equation \ref{spectralage} relates the source age and lobe pressure as $t \propto p^{-0.75}$, where $B \gg B_{\rm ic}(z)$. This assumption is justified since all 37 3C sources fitted using our {\sc Full\,Fit} algorithm have lobe magnetic field strengths greater than the cosmic microwave background equivalent, typically by at least a factor of a few. By contrast, the luminosity constraint depends on the age and pressure only through the $p(t)^{(s + 5)/4} V(t)$ term (Equation \ref{luminosityloss}), where in the strong-shock expansion limit $V(t) \propto t^{9/(5 - \beta)}$ \citep{KA+1997}. The pressure and age are therefore related as $t \propto p^{(s + 5)(\beta - 5)/36}$ in this constraint. For typical values of $\beta = 1.5$ and $s = 2.5$ this relationship becomes $t \propto p^{-0.73}$, mirroring that of the spectral break frequency constraint. The source age and lobe pressure can therefore not be distinguished solely by the luminosity and break frequency constraints.

The degeneracy between the age and lobe pressure has the unexpected benefit that these parameters can effectively be treated as a single variable, reducing the number of unknown parameters in the fitting algorithm. The jet power and equipartition factor parameters can thus be determined accurately with one less observable than required in the full model. The {\sc Spectral Break} fitting algorithm should therefore be capable of estimating jet powers and equipartition factors with comparable accuracy to the {\sc Full\,Fit} model, but have no ability to constrain the source age. 

\subsection{Jet power--equipartition factor degeneracy}

The size constraint relates the source age and jet power in the strong-shock supersonic limit as $D \propto t^{3/(5 - \beta)} Q^{1/(5 - \beta)}$ \citep[Equation 4 of][]{KA+1997}. The much stronger dependence on the source age parameter places a tighter constraint on this parameter in an underdetermined fitting algorithm, with good source age estimates expected for only order of magnitude accurate estimates of jet powers. Source ages estimated using the {\sc Source\,Size} algorithm should therefore have close to comparable accuracy to the {\sc Full\,Fit} algorithm; we discuss this further in Section \ref{sec:Size--luminosity fitting algorithm}.

By contrast, the luminosity constraint relates these parameters and the equipartition factor in the strong-shock limit as

\begin{equation}
\begin{split}
L = L_0 &t^{[36 - (s + 5)(4 + \beta)]/4(5 - \beta)} \\
&\quad Q^{[12 + (s + 5)(2 - \beta)]/4(5 - \beta)} q^{(s + 1)/4} ,
\end{split}
\end{equation}

where $L_0$ is a constant of proportionality including the loss processes and lobe plasma composition, and again $q = (B/B_{\rm eq})^{(s + 5)/2}$. For typical values of $\beta = 1.5$ and $s = 2.5$, this relationship becomes $L \propto t^{-0.38} Q^{1.13} q^{0.875}$, or for a given best-fit source age is $L \propto Q^{1.25} q^{0.875}$. The similar exponents on $Q$ and $q$ mean neither the jet power nor equipartition factor can be well constrained. Typical values for one of these parameters must be assumed in order to better constrain the other. In this work we assume a prior probability distribution for the equipartition factor (see Section \ref{sec:Lobe equipartition factor} and Figure \ref{fig:equidist_plot}) to enable the jet power to be well constrained in the {\sc Source\,Size} algorithm; {note that this prior is not applied in Figures \ref{fig:QQplot}, \ref{fig:ttplot} and \ref{fig:eqeqplot}}.

\section{Fitted radio source properties}

\begin{table*}
\begin{center}
\caption[]{Radio source jet powers, ages and equipartition factors estimated for our sample of 37 3C radio AGN with the requisite data to be fitted using the {\sc Full\,Fit} algorithm. The redshift, size and luminosity observables are extracted from \citet{Mullin+2008}, the stellar masses are calculated using flux densities taken from \citet{Willott+2003} and \citet{Mullin+2008}, and the spectral break frequencies are derived from fits to the multi-frequency luminosity observations of \citet{Laing+1980}. {The goodness of fit of the CI model to their spectra across the entire 0.01-$10\rm\, GHz$ frequency range is included in the final column (i.e. probability that the CI model fits the data); spectra which are well fitted with the removal of a single outlying low-frequency point are marked with a dagger and those with multiple inconsistent points with stars. The break frequencies of these noisy spectra are well fitted by supplementing the \citet{Laing+1980} observations with additional literature data.} Pictor A is calculated using observations by \citet{Kuehr+1981} and \citet{Hardcastle+2016}; some caution should be exercised when comparing these results due to the use of different datasets.}
\label{tab:fittedvalues}
\setlength{\tabcolsep}{10pt}
\begin{tabular}{ccccccccccc}
\hline\hline \vspace{-0.00cm}
Source&$z$&$M_\star$&D&$L_{178}$&$\nu_{\rm b}$&$Q$&$t_{\rm age}$&$B/B_{\rm eq}$&CI Fit \\
&&\!\![log $\rm M_\odot$]\!\!&[kpc]&\!\!\![log $\rm W\!/Hz$]\!\!\!&[log Hz]&[log W]&\![log yrs]\!&&\!\!\![p-value]\!\!\!\vspace{0.000cm}\\
\hline \vspace{-0.00cm}
3C6.1&0.8404&-&207$\pm$8&28.62&9.5$\pm$0.6&39.6$\pm$0.2&7.3$\pm$0.2&0.34$\pm$0.04&0.93\\
3C19&0.482&11.89&40$\pm$1&28.00&9.8$\pm$0.6&38.6$\pm$0.2&6.8$\pm$0.3&0.36$\pm$0.04&0.88\\
3C20&0.174&11.52&151$\pm$2&27.57&9.7$\pm$1.2&38.4$\pm$0.2&7.5$\pm$0.3&0.43$\pm$0.06&0.91\\
3C41&0.795&11.62&185$\pm$7&28.41&9.9$\pm$0.5&39.7$\pm$0.2&7.2$\pm$0.3&0.27$\pm$0.03&0.59\\
3C42&0.395&11.73&150$\pm$1&27.82&9.7$\pm$0.7&38.3$\pm$0.2&7.7$\pm$0.2&0.34$\pm$0.04&0.76\\
3C55&0.735&11.75&510$\pm$7&28.77&8.1$\pm$1.3&39.0$\pm$0.2&8.2$\pm$0.2&0.55$\pm$0.09&0.19\\
3C67&0.3102&-&15$\pm$2&27.48&9.8$\pm$1.0&38.0$\pm$0.2&6.4$\pm$0.3&0.43$\pm$0.05&0.10\\
3C79&0.2559&11.66&366$\pm$23&27.82&9.0$\pm$1.0&38.0$\pm$0.2&8.3$\pm$0.2&0.43$\pm$0.05&0.17\\
3C123&0.2177&11.73&110$\pm$7&28.43&10.0$\pm$0.4&38.9$\pm$0.2&7.4$\pm$0.3&0.28$\pm$0.03&<0.05$^\dagger$\\
3C132&0.214&11.66&77$\pm$1&27.27&10.0$\pm$0.6&38.3$\pm$0.2&7.0$\pm$0.2&0.44$\pm$0.05&<0.05**\\
3C153&0.2769&11.80&34$\pm$3&27.57&9.3$\pm$1.0&38.2$\pm$0.2&6.8$\pm$0.3&0.51$\pm$0.07&<0.05$^\dagger$\\
3C171&0.2384&11.49&38$\pm$0&27.55&10.0$\pm$0.6&37.8$\pm$0.2&7.2$\pm$0.3&0.35$\pm$0.04&0.69\\
3C175.1&0.92&11.58&61$\pm$3&28.70&9.4$\pm$0.7&39.5$\pm$0.2&6.7$\pm$0.3&0.34$\pm$0.04&0.58\\
3C184&0.994&-&38$\pm$5&28.83&9.0$\pm$0.5&39.3$\pm$0.2&6.6$\pm$0.3&0.38$\pm$0.05&0.68\\
3C196&0.871&-&48$\pm$4&29.38&9.0$\pm$1.2&39.6$\pm$0.2&6.8$\pm$0.3&0.31$\pm$0.04&0.64\\
3C217&0.8975&11.48&104$\pm$24&28.63&9.0$\pm$0.8&39.6$\pm$0.2&6.7$\pm$0.3&0.53$\pm$0.07&<0.05**\\
3C219&0.1744&11.77&566$\pm$8&27.57&9.6$\pm$1.2&38.1$\pm$0.3&8.5$\pm$0.2&0.32$\pm$0.04&0.99\\
3C220.3&0.685&-&70$\pm$5&28.49&8.8$\pm$0.4&38.6$\pm$0.2&7.2$\pm$0.3&0.49$\pm$0.06&<0.05$^\dagger$\\
3C226&0.82&11.34&301$\pm$16&28.69&9.5$\pm$0.8&39.4$\pm$0.2&7.6$\pm$0.2&0.35$\pm$0.04&0.76\\
3C244.1&0.428&-&289$\pm$7&28.14&9.9$\pm$0.5&39.5$\pm$0.2&7.3$\pm$0.2&0.42$\pm$0.05&0.97\\
3C247&0.7489&11.99&104$\pm$9&28.38&9.4$\pm$1.1&38.7$\pm$0.2&7.6$\pm$0.3&0.29$\pm$0.04&0.27\\
3C263.1&0.824&11.82&67$\pm$13&28.77&8.9$\pm$1.0&39.4$\pm$0.2&6.8$\pm$0.3&0.46$\pm$0.06&0.99\\
3C268.1&0.9731&-&339$\pm$19&28.93&9.8$\pm$0.6&39.8$\pm$0.2&7.9$\pm$0.2&0.21$\pm$0.03&0.72\\
3C280&0.996&11.90&135$\pm$8&29.07&10.2$\pm$0.4&40.0$\pm$0.2&7.1$\pm$0.3&0.20$\pm$0.02&0.97\\
3C289&0.9674&11.91&84$\pm$1&28.74&9.1$\pm$0.9&39.5$\pm$0.2&6.9$\pm$0.3&0.38$\pm$0.05&0.93\\
3C300&0.272&11.43&402$\pm$84&27.63&9.7$\pm$0.6&38.6$\pm$0.2&7.8$\pm$0.2&0.47$\pm$0.06&0.99\\
3C319&0.192&11.25&340$\pm$14&27.24&8.2$\pm$1.5&37.2$\pm$0.2&8.6$\pm$0.2&0.62$\pm$0.11&0.98\\
3C325&0.86&12.18&140$\pm$11&28.71&9.1$\pm$0.6&39.8$\pm$0.2&7.0$\pm$0.3&0.40$\pm$0.05&0.07\\
3C330&0.549&11.86&399$\pm$3&28.51&9.7$\pm$0.5&39.6$\pm$0.2&7.6$\pm$0.2&0.36$\pm$0.04&0.23\\
3C340&0.7754&11.64&327$\pm$10&28.42&9.6$\pm$0.8&39.4$\pm$0.2&7.5$\pm$0.2&0.36$\pm$0.04&0.76\\
3C352&0.806&11.75&101$\pm$3&28.55&8.4$\pm$0.3&38.7$\pm$0.2&7.3$\pm$0.3&0.55$\pm$0.09&<0.05**\\
3C388&0.0908&11.85&85$\pm$5&26.73&9.7$\pm$1.0&37.1$\pm$0.1&8.1$\pm$0.2&0.23$\pm$0.03&0.87\\
3C401&0.201&-&80$\pm$3&27.40&9.0$\pm$0.5&37.9$\pm$0.2&7.4$\pm$0.2&0.55$\pm$0.09&0.77\\
3C427.1&0.572&11.47&178$\pm$7&28.58&8.3$\pm$1.3&39.1$\pm$0.2&7.3$\pm$0.2&0.72$\pm$0.13&0.99\\
3C438&0.29&11.96&96$\pm$3&28.10&9.0$\pm$0.3&38.0$\pm$0.2&7.6$\pm$0.2&0.39$\pm$0.05&<0.05$^\dagger$\\
3C441&0.708&11.85&261$\pm$52&28.45&10.1$\pm$0.6&39.6$\pm$0.2&7.4$\pm$0.2&0.27$\pm$0.03&0.55\\
3C455&0.5427&-&27$\pm$4&28.16&9.2$\pm$0.7&38.4$\pm$0.2&6.7$\pm$0.3&0.43$\pm$0.05&0.75\\
PictorA&0.03506&-&56$\pm$2&27.04&8.8$\pm$1.1&37.0$\pm$0.2&7.6$\pm$0.2&0.51$\pm$0.07&-\vspace{0.000cm}\\
\hline
\end{tabular}
\end{center}
\end{table*}

\begin{table*}
\begin{center}
\caption[]{Radio source jet powers and ages estimated for our sample of 34 3C radio AGN with only sufficient data to be fitted using the {\sc Source\,Size} algorithm. {These results are calculated assuming a prior probability distribution for the equipartition factor.} The redshift, size and luminosity observables are extracted from \citet{Mullin+2008}, and the stellar masses are calculated using flux densities taken from \citet{Willott+2003} and \citet{Mullin+2008}. {The goodness of fit of the CI model across the 0.01-$10\rm\, GHz$ frequency range is included in the final column (see Table \ref{tab:fittedvalues}).}}
\label{tab:fittedvalues2}
\setlength{\tabcolsep}{10pt}
\begin{tabular}{cccccccccc}
\hline\hline \vspace{-0.00cm}
Source&$z$&$M_\star$&D&$L_{178}$&$Q$&$t_{\rm age}$&CI Fit \\
&&\!\![log $\rm M_\odot$]\!\!&[kpc]&\!\!\![log $\rm W\!/Hz$]\!\!\!&[log W]&\![log yrs]\!&\!\!\![p-value]\!\!\!\vspace{0.000cm}\\
\hline \vspace{-0.00cm}
3C16&0.405&11.45&413$\pm$78&27.84&39.6$\pm$0.2&7.6$\pm$0.2&0.29\\
3C22&0.938&11.73&206$\pm$7&28.71&40.1$\pm$0.2&7.1$\pm$0.2&0.86\\
3C33&0.0595&11.59&292$\pm$13&26.70&38.7$\pm$0.2&7.6$\pm$0.2&0.24\\
3C33.1&0.181&-&707$\pm$99&27.09&38.4$\pm$0.3&8.4$\pm$0.2&0.94\\
3C34&0.689&11.83&327$\pm$2&28.45&39.9$\pm$0.2&7.4$\pm$0.2&0.84\\
3C35&0.0677&11.68&905$\pm$5&26.10&37.9$\pm$0.3&9.0$\pm$0.1&0.14\\
3C46&0.4373&11.95&978$\pm$99&27.91&39.9$\pm$0.2&7.9$\pm$0.2&0.93\\
3C47&0.425&-&424$\pm$17&28.27&39.6$\pm$0.2&7.8$\pm$0.2&0.40\\
3C61.1&0.186&11.54&563$\pm$46&27.50&39.4$\pm$0.2&7.7$\pm$0.2&<0.05$^\dagger$\\
3C98&0.0306&11.40&169$\pm$1&26.04&36.9$\pm$0.1&8.8$\pm$0.4&<0.05**\\
3C172&0.5191&11.78&584$\pm$7&28.21&39.9$\pm$0.2&7.7$\pm$0.2&0.58\\
3C173.1&0.292&11.88&261$\pm$11&27.64&38.9$\pm$0.2&7.7$\pm$0.2&0.40\\
3C175&0.768&-&392$\pm$36&28.71&40.2$\pm$0.2&7.4$\pm$0.2&0.86\\
3C184.1&0.1187&11.37&395$\pm$29&26.70&38.9$\pm$0.2&7.6$\pm$0.2&0.12\\
3C192&0.0598&11.43&234$\pm$10&26.29&38.4$\pm$0.2&7.5$\pm$0.2&0.28\\
3C200&0.458&11.69&147$\pm$9&27.95&39.2$\pm$0.2&7.3$\pm$0.3&0.99\\
3C223&0.1368&11.45&738$\pm$5&26.89&39.0$\pm$0.2&8.0$\pm$0.2&0.95\\
3C225B&0.58&11.62&33$\pm$3&28.49&39.5$\pm$0.2&6.3$\pm$0.3&0.97\\
3C228&0.5524&11.60&307$\pm$10&28.46&40.0$\pm$0.2&7.3$\pm$0.2&0.18\\
3C234&0.1848&11.83&295$\pm$52&27.51&38.2$\pm$0.3&8.3$\pm$0.2&1.00\\
3C249.1&0.311&-&222$\pm$25&27.54&38.8$\pm$0.2&7.7$\pm$0.2&0.67\\
3C265&0.8108&12.06&585$\pm$58&28.81&39.6$\pm$0.2&8.3$\pm$0.1&0.99\\
3C274.1&0.422&11.71&921$\pm$31&28.04&40.1$\pm$0.2&7.7$\pm$0.2&0.95\\
3C277.2&0.766&11.61&439$\pm$96&28.55&40.4$\pm$0.2&7.3$\pm$0.2&0.83\\
3C284&0.2394&11.79&701$\pm$65&27.32&38.9$\pm$0.3&8.2$\pm$0.2&0.74\\
3C285&0.0794&11.55&265$\pm$13&26.28&37.2$\pm$0.1&8.9$\pm$0.3&0.65\\
3C303&0.141&11.70&121$\pm$9&26.80&37.3$\pm$0.2&8.4$\pm$0.3&<0.05**\\
3C336&0.927&-&201$\pm$19&28.66&39.9$\pm$0.2&7.2$\pm$0.2&0.76\\
3C341&0.448&11.77&449$\pm$34&27.88&39.9$\pm$0.2&7.4$\pm$0.2&0.99\\
3C337&0.635&11.62&307$\pm$44&28.26&39.7$\pm$0.2&7.5$\pm$0.2&0.10\\
3C349&0.205&11.48&287$\pm$7&27.23&39.0$\pm$0.2&7.6$\pm$0.2&0.27\\
3C381&0.1605&11.63&202$\pm$8&27.09&38.6$\pm$0.2&7.6$\pm$0.3&0.30\\
3C436&0.2145&11.76&375$\pm$14&27.40&38.8$\pm$0.2&7.9$\pm$0.2&0.95\\
3C452&0.0811&11.73&428$\pm$5&26.98&37.7$\pm$0.3&8.5$\pm$0.2&0.72\vspace{0.000cm}\\\hline
\end{tabular}
\end{center}
\end{table*}

\end{appendix}

\end{document}